\documentclass[journal]{IEEEtran}

\usepackage{xcolor}
\usepackage{graphicx}
\usepackage{amsmath, bbm}
\usepackage{amssymb}
\usepackage{booktabs}
\usepackage{multirow}
\usepackage{array,multirow}
\usepackage{cite}
\usepackage{textcomp}
\usepackage{algorithm,algorithmic}
\usepackage{fancyhdr}

\begin{document}

\title{Deep Frame Prediction for Video Coding}

\author{Hyomin~Choi,~\IEEEmembership{Student Member,~IEEE,}
        and~Ivan~V.~Baji\'{c},~\IEEEmembership{Senior Member,~IEEE}
\thanks{H. Choi and I. V. Baji\'{c} are with the School of Engineering Science, Simon Fraser University, BC, V5A 1S6, Canada. E-mail: chyomin@sfu.ca, ibajic@ensc.sfu.ca. This work was supported in part by the Vanier Scholarship and the NSERC Grant RGPIN-2016-04590.}}

\maketitle

\thispagestyle{empty}
\renewcommand{\headrulewidth}{0.0pt}
\thispagestyle{fancy}
\lhead{}
\chead{Copyright \copyright 2019 IEEE. Personal use of this material is permitted. However, permission to use this material for any other purposes must be obtained from the IEEE by sending an email to pubs-permissions@ieee.org.}
\rhead{}
\lfoot{}
\cfoot{}
\rfoot{}

\begin{abstract}
We propose a novel frame prediction method using a deep neural network (DNN), with the goal of improving video coding efficiency. The proposed DNN makes use of decoded frames, at both encoder and decoder, to predict textures of the current coding block. Unlike conventional inter-prediction, the proposed method does not require any motion information to be transferred between the encoder and the decoder.  
Still, both uni-directional and bi-directional prediction are possible using the proposed DNN, which is enabled by the use of the temporal index channel, in addition to color channels. 
In this study, we developed a jointly trained DNN for both uni- and bi-directional prediction, as well as separate networks for uni- and bi-directional prediction, and compared the efficacy of both approaches. The proposed DNNs were  compared with the conventional motion-compensated prediction in the latest video coding standard, HEVC, in terms of BD-Bitrate. The experiments show that the proposed joint DNN (for both uni- and bi-directional prediction) reduces the luminance bitrate by about 4.4\%, 2.4\%, and 2.3\% in the Low delay P, Low delay, and Random access configurations, respectively. In addition, using the separately trained DNNs brings further bit savings of about 0.3\%--0.5\%.
\end{abstract}


\IEEEpeerreviewmaketitle

\section{Introduction}
\IEEEPARstart{W}{ith} increasing demand for video services~\cite{CISCO_VNI_2017, Surveillance_video}, the need for more efficient video coding is also growing. In response to such demand, Joint Video Experts Team (JVET) was established after the latest video coding standard, High Efficiency Video Coding (HEVC), and has been exploring future video coding technologies. In 2018, JVET called for proposals for the next video coding standard, tentatively named Versatile Video Coding (VVC). Some of the submitted proposals considerably surpass the performance of HEVC in terms of both subjective and objective qualities~\cite{jvet_j1003}. Interestingly, several contributions proposed Deep Neural Network (DNN)-aided coding tools. 

Recently, DNN-based coding tools have become a topic of interest, and experimental results superior to the conventional coding tools in terms of rate-distortion (RD) performance have started to appear in the literature. Some of these have also been discussed in JVET meetings as part of normative tools~\cite{ahg9_macao}. For example, DNN-based quality improvement for in-loop filtering and post-filtering have been actively researched in both academia and the standardization community~\cite{yang2017decoder, jin2018quality, li2017cnn, he2018enhancing, kang2017multi, jia2017spatial, in_loop_for_intra, in_loop_for_inter, jvet_k0158, jvet_k0222,jvet_k0391, jvet_l0242, jvet_l0383}. Other aspects of video coding where DNNs have been suggested are 
intra and inter prediction \cite{cui2017convolutional, li2017intra, hu2018enhanced, hu2018progressive, JVET_J0037_sandiego_intra_hhi, zhang2017learning, yan2018convolutional, huo2018convolutional, zhao2018cnn, zhao2018enhanced, hevc_with_sep_conv_for_ra}, reduction of coding complexity~\cite{jin2017cnn, xu2018reducing, wang2018fast}, modeling the RD relationship~\cite{xu2017cnn}, and replacing the entire coding pipeline by a DNN~\cite{chen2017deepcoder}. Given the variety of coding tools where DNNs have already made inroads, they appear to be an inevitable technological trend in video compression.

For intra prediction,~\cite{cui2017convolutional, hu2018progressive} employed a Convolutional Neural Network (CNN) with neighbouring blocks (instead of reference lines) as inputs, to generate a predicted block. Moreover,~\cite{hu2018enhanced} included a Recurrent Neural Network (RNN) into their predictor for improved accuracy. A fully connected network architecture (Multi-Layer Perceptron, MLP) is adopted in~\cite{li2017intra, JVET_J0037_sandiego_intra_hhi} for intra prediction, and the input samples are also from the neighboring region. Specifically, ~\cite{JVET_J0037_sandiego_intra_hhi} proposed a Discrete Cosine Transform (DCT)-domain predictor, where the output of the MLP is subject to inverse DCT to obtain the predicted pixels.

However, most of the coding gain in video  compression comes from inter prediction. Recent works on inter prediction using DNNs include interpolation filters~\cite{zhang2017learning, yan2018convolutional}, enhanced motion compensation~\cite{huo2018convolutional, zhao2018cnn, zhao2018enhanced}, and texture prediction, which directly generates the predicted pixel values using reference samples/frames as inputs to the trained network~\cite{hevc_with_sep_conv_for_ra} without transferring motion information. In this paper, we propose a DNN-based frame prediction architecture that is able to support both uni- and bi-directional prediction. Our work is inspired by recent progress on frame prediction (Section~\ref{sec:literature_review}), and is the first, to our knowledge, to develop a single DNN that can support both P and B frame coding. 

The paper is organized as follows. Section~\ref{sec:literature_review} reviews recent related work on DNN-based inter prediction, and identifies the contributions of this paper. In Section~\ref{sec:proposed_method}, presents the architecture of the proposed DNN and describes how the proposed DNN-based prediction can be used in video coding. Section~\ref{sec:network_training} describes the training strategy and presents an ablation study for the proposed DNN.  Section~\ref{sec:experimental_results} presents the experimental results of prediction performance and compression efficiency of the proposed DNN in various configurations. Finally, the paper is concluded in Section~\ref{sec:conclusion}.

\begin{figure*}[t]
    \begin{minipage}[b]{1.0\linewidth}
    \centering
    \includegraphics[width=\textwidth]{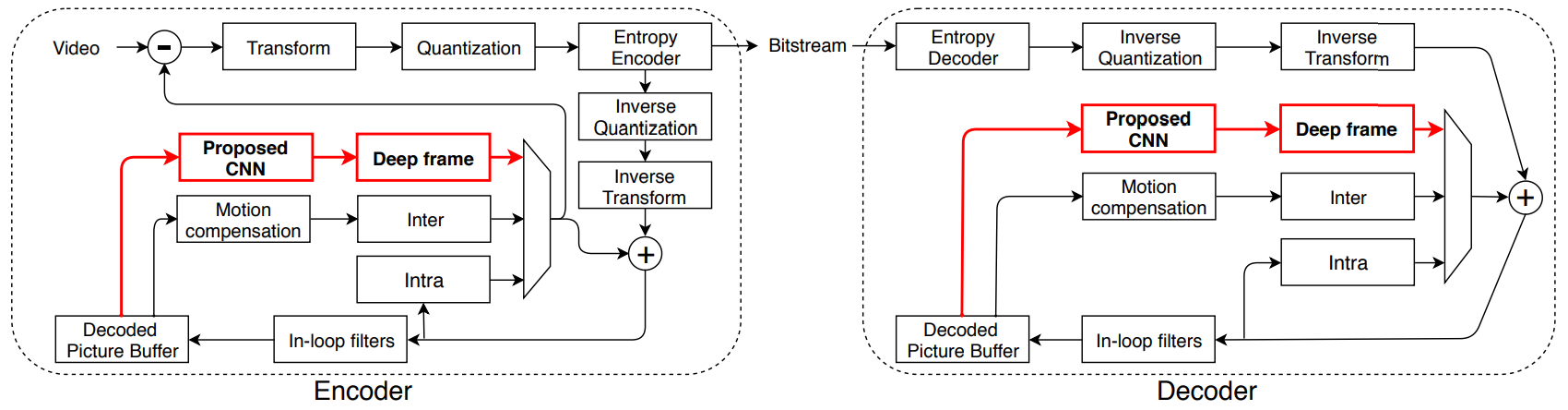}
    \end{minipage}
\vspace{-0.6cm}
\caption{The proposed deep frame prediction in the encoder (left) and decoder (right).}
\label{fig:decoder_with_proposed_method}
\vspace{-.3cm}
\end{figure*}

\section{Neural network-aided inter prediction}
\label{sec:literature_review}
\subsection{Prior work}
Inter coding plays a crucial role in achieving high efficiency in video compression. Fractional-pel motion compensation requires a frame to be interpolated in between existing pixels to compensate for continuous motion of objects. In HEVC, DCT-based interpolation filters  provide quarter-pel precision, but their coefficients are non-adaptive, which may limit their effectiveness~\cite{ugur2013motion}. In~\cite{zhang2017learning, yan2018convolutional}, CNNs have been used for content-adaptive interpolation. Specifically, Zhang \textit{et al.}~\cite{zhang2017learning} proposed a half-pel interpolation filter based on a super-resolution network~\cite{kim2016accurate}, while Yan \textit{et al.}~\cite{yan2018convolutional}  proposed a different CNN-based interpolation filter. 

Other studies that have looked at the use of neural networks in motion compensation include~\cite{huo2018convolutional, zhao2018cnn, zhao2018enhanced}. Huo \textit{et al.}~\cite{huo2018convolutional} proposed a CNN-based motion compensation refinement algorithm. The suggested network has the motion-compensated block and its neighboring coded area as inputs, and it generates the refined prediction block. Considering the spatial correlation between adjacent pixels, especially for small blocks, this approach is able to reduce some artifacts along the block boundaries. The approach is less suitable for larger blocks, but it is applicable to both  uni- and bi-prediction. In~\cite{zhao2018cnn, zhao2018enhanced}, Zhao \textit{et al.} suggested CNN-based bi-directional motion compensation. Conventional bi-prediction averages two predictions 
unless additional weights are transmitted, while the proposed CNN  combines two predictions adaptively, based on the content, and produces more accurate predicted blocks. Nonetheless, this method is only applicable to bi-directional prediction. 

Prior work that is most relevant to the present study is~\cite{hevc_with_sep_conv_for_ra}, where Zhao \textit{et al.} proposed DNN-aided bi-directional prediction in HEVC-based video coding, and reported considerable coding gains for high quantization parameter (QP) values in the Random Access (RA) configuration. However, this work merely adopted the frame rate up conversion (FRUC) network~\cite{Niklaus_ICCV_2017} for the purpose of video compression, even using the weights obtained in~\cite{Niklaus_ICCV_2017}. Due to these reasons, the approach in~\cite{hevc_with_sep_conv_for_ra} is limited by the FRUC constraints imposed in~\cite{Niklaus_ICCV_2017}, as follows: (1) it does not properly synthesize the intermediate target frame unless the reference frames are symmetrically spaced around the target frame, and (2) it cannot predict a future frame (uni-directional prediction) using previous frames, for low-delay coding scenarios.

\subsection{Our contribution}
In this paper, we propose solutions to the issues listed above. Specifically, the contributions of the present work are as follows:
\begin{itemize}
    \item We develop a DNN that can perform both uni- and bi-directional prediction.
    \item The proposed DNN is able to handle frames at various distances from the target frame.
    \item The proposed DNN is able to operate in all inter coding scenarios and provides a unified frame prediction tool for video coding.
    \item We describe the integration of the proposed DNN into HM reference encoder and decoder.
    \item We present an ablation study identifying the contribution of individual components of the proposed DNN.
    \item We analyze the video bitstream generated by using the proposed frame prediction in video coding and identify where the savings are being made compared with conventional HEVC.
\end{itemize}

\section{Proposed frame prediction}
\label{sec:proposed_method}

The proposed method is an additional tool for video compression that supplements traditional intra/inter prediction, as shown in Fig.~\ref{fig:decoder_with_proposed_method}. The proposed CNN requires two frames from the decoded picture buffer (DPB), along with their temporal indices, as inputs. The network produces filter coefficients that are used to synthesize patches of a new frame, which we refer to as \emph{deep frame}. Once the deep frame is synthesized, it is used directly as a predictor for the current frame. Therefore, with this form of inter-prediction, no additional motion information needs to be coded, only the prediction residual and block-based flags indicating the use of the proposed method. As shown in Fig.~\ref{fig:decoder_with_proposed_method}, the same procedure is performed at the decoder to synthesize the deep frame, and then the residual is added to produce the final decoded frame.

\subsection{Proposed network architecture}
\label{ssec:the_neural_net_structure}

\begin{figure*}[t]
    \begin{minipage}[b]{1.0\linewidth}
    \centering
    \includegraphics[width=\textwidth]{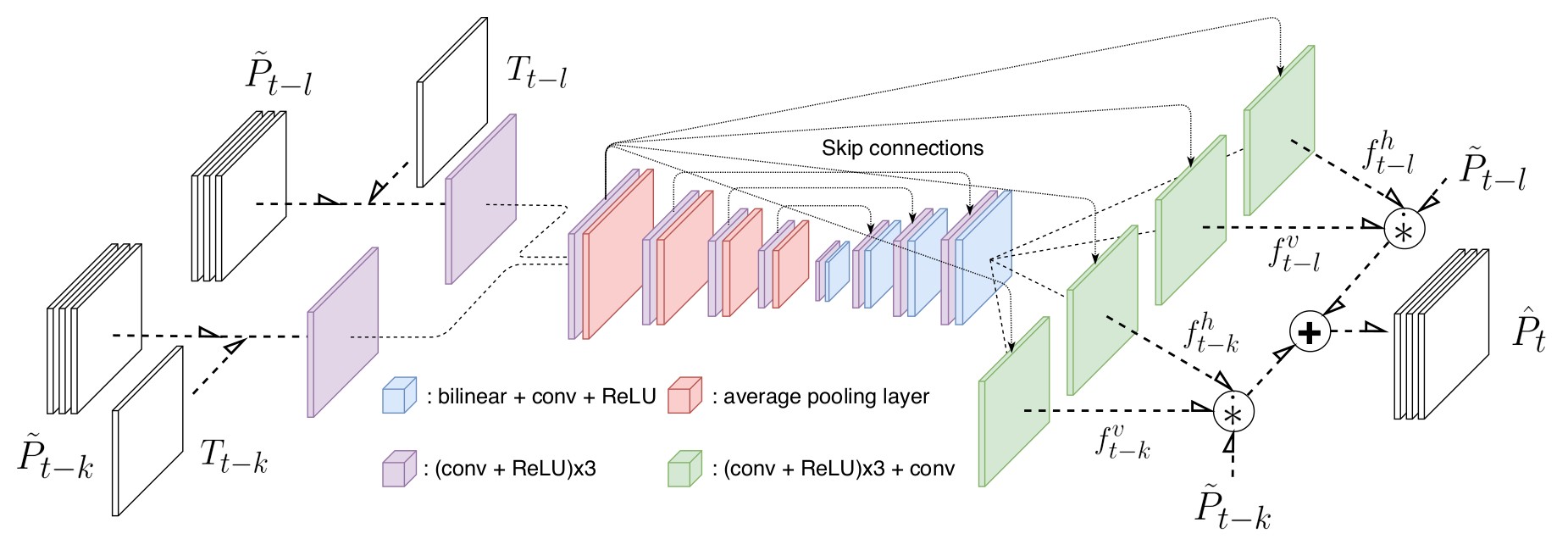}
    \end{minipage}
\caption{Architecture of the proposed neural network. Four-channel tensors derived from two reference patches and their temporal indices are applied at the two inputs. Processing is performed in ten blocks whose structure is indicated in the figure and whose input/output dimensions are shown in Table~\ref{tbl:layer_structure}. The outputs produce spatially-varying filters that are used to synthesize the predicted patch.}
\label{fig:network_architecture}
\end{figure*}

\begin{table*}[t]
\centering
\caption{Input/output dimensions of various blocks (B1--B10) in the proposed network}
\label{tbl:layer_structure}
\resizebox{\linewidth}{!}{%
\setlength{\tabcolsep}{2pt}
\begin{tabular}{@{}c|c|c|c|c|c|c|c|c|c|c@{}}
\toprule
       & \begin{tabular}[c]{@{}c@{}}B1 \\ (two paths)\end{tabular} & \begin{tabular}[c]{@{}c@{}}B2\end{tabular} & \begin{tabular}[c]{@{}c@{}}B3\end{tabular} & \begin{tabular}[c]{@{}c@{}}B4\end{tabular} & \begin{tabular}[c]{@{}c@{}}B5\end{tabular} & \begin{tabular}[c]{@{}c@{}}B6\end{tabular} & \begin{tabular}[c]{@{}c@{}}B7\end{tabular} & \begin{tabular}[c]{@{}c@{}}B8\end{tabular} & \begin{tabular}[c]{@{}c@{}}B9\end{tabular} & \begin{tabular}[c]{@{}c@{}}B10 \\ (four paths) \end{tabular} \\ \midrule
Input  & $N \times M \times 4$                                                 & $N \times M \times 32$                                                            & $\frac{N}{2} \times \frac{M}{2} \times 64$                                                            & $\frac{N}{4} \times \frac{M}{4} \times 128$                                                            & $\frac{N}{8} \times \frac{M}{8} \times 256$                                                            & $\frac{N}{16} \times \frac{M}{16} \times 512$              & $\frac{N}{8} \times \frac{M}{8} \times 512$                                                             & $\frac{N}{4} \times \frac{M}{4} \times 256$                                                             & $\frac{N}{2} \times \frac{M}{2} \times 128$                                                             & $N \times M \times 64$                                                     \\[0.15cm]
Output & $N \times M \times 16$                                                 & $\frac{N}{2} \times \frac{M}{2} \times 64$                                                       & $\frac{N}{4} \times \frac{M}{4} \times 128$                                                  & $\frac{N}{8} \times \frac{M}{8} \times 256$                                                  & $\frac{N}{16} \times \frac{M}{16} \times 512$                                                & $\frac{N}{8} \times \frac{M}{8} \times 512$ & $\frac{N}{4} \times \frac{M}{4} \times 256$                                                   & $\frac{N}{2} \times \frac{M}{2} \times 128$                                                   & $N \times M \times 64$                                                    & $N \times M \times C$                                                 \\ \bottomrule
\end{tabular}}
\end{table*}

Our DNN architecture for  deep frame prediction is shown in Fig.~\ref{fig:network_architecture}. It is inspired by the work of Niklaus \textit{et al.}~\cite{Niklaus_ICCV_2017}, who proposed a similar network for FRUC. Their network creates a new frame by interpolating mid-way between two consecutive frames. However, in addition to bi-directional interpolation, our proposed DNN (Fig.~\ref{fig:network_architecture}) also supports uni-directional prediction (i.e., extrapolation), and is able to handle reference frames at varying distances from the predicted frame.
In video coding, it is important to be able to support both uni- and bi-directional prediction because, depending on the settings for different coding scenarios, some frames might not be available for prediction. This is true in low-delay cases, and more generally, in cases where the coding order does not produce an available decoded frame on both sides of the currently coded frame. 

The proposed network has two input paths, which are then merged inside the network. The two inputs are fed with two patch tensors ($\widetilde{\mathbf{P}}_{t_1}$ and $\widetilde{\mathbf{P}}_{t_2}$) of size $N \times M \times 3$ (three color channels), from which prediction of an $N \times M \times 3$ patch in the current frame is made. For training the network we used $N=M=128$, but in testing, different values are used depending on the resolution of the coded sequence. 
Since the proposed DNN does not contain any fully-connected layers, the output scales with input size. Indices $t_1$ and $t_2$ (with $t_1 < t_2$) represent time index relative to the current frame index $t$. The tilde character (\texttildelow) indicates that these patches come from previously decoded frames. If needed, the patches are converted from YUV420 to YUV444 to make the resolution of all color channels the same, so that conventional convolutional layers can process the input. The final frame is converted back to YUV420.

In addition to the color channels, the input tensor contains an additional temporal index channel. As will be seen in the ablation study in Section~\ref{ssec:pre-training and ablation study}, this channel makes a significant contribution to the performance of the proposed network. Specifically, to the tensor $\widetilde{\mathbf{P}}_{t_i}, i \in \{1,2\}$, we append $\mathbf{T}_{t_i} = c_i\cdot\mathbf{1}_{N \times M}$, where $\mathbf{1}_{N \times M}$ is a $N \times M$ matrix of all ones, and constants $c_i$ are chosen as
\begin{equation}
\centering
(c_1, c_2) = 
    \begin{cases}
    (-10, 10), & \text{if } t_1 < t < t_2,\\
    (-20, -10), & \text{if } t_1 < t_2 < t. 
    \end{cases}
\label{eq:temporal_scale}
\end{equation}
The sign of $c_i$ indicates whether the corresponding patch comes from a previous or subsequent frame, and its magnitude indicates the relative distance to the current frame. We also experimented with making $c_i$ a multiple of the scaled frame distance, but found that the best results are achieved when the reference frames are closest to the frame being predicted, and those cases can be handled by the values in~(\ref{eq:temporal_scale}).
Using indices as tensor channels is inspired by the work in~\cite{liu2018intriguing_UBER_COORDNET} where the authors have used spatial coordinates as additional tensor channels to achieve  spatially-variant processing. Here, we use this concept for temporal indices of reference frames to enable temporally-variant processing that depends on the relative positions of the reference frames and the frame to be synthesized. 

Processing within the network is accomplished via various processing blocks (B1, B2, ..., B10), which are indicated in Fig.~\ref{fig:network_architecture}. In the figure legend, the following abbreviations are used: ``conv'' stands for the convolutional layer (with $3 \times 3$ filters), ``ReLU'' stands for Rectified Linear Unit activation, and ``bilinear'' stands for bilinear upsampling. Input/output dimensions of various processing blocks are indicated in Table~\ref{tbl:layer_structure}. As the typical design choice, the number of output channels for the inner blocks is a power of 2, and each spatial dimension changes between consecutive blocks in a multiple of 2 or a multiple of 1/2, depending on whether downsampling or upsampling is performed. The number of output channels for the last block B10 is $C$, which determines the size of the 2-D prediction kernels, as explained later in this section. We used $C=51$ in the experiments. To avoid data inflation at the final stage, the number of input channels to B10 should be more than $C$, so we chose 64, which is the smallest power of 2 larger than $C=51$. Working backwards from there, we determined the number of channels for each block, as shown in Table~\ref{tbl:layer_structure}.

The two inputs are initially processed separately by the adaptation block B1, whose purpose is to fuse the spatial information (pixels in each frame) and temporal information (temporal index channel). Its impact will be assessed in the ablation study in Section~\ref{ssec:pre-training and ablation study}. Subsequently, the data from the two input paths is merged and fed to the inner portion of our DNN, which resembles a U-Net~\cite{Ronneberger2015UNetCN}. 
We used a depth-4 U-Net with skip connections. The depth was chosen so that the data volume at the bottleneck is smaller, but still similar, to the input data volume. Recall, the input to the DNN are two color (3-channel) frames of size $M\times N$, each with its own temporal index channel, for a total data volume of $8\cdot M \cdot N$. With the selected depth-4 U-Net, the data volume at the bottleneck is $\frac{M}{16} \times \frac{N}{16} \times 512 = 2 \cdot M \cdot N$. 
Another modification to the original U-Net is the addition of skip connections from the merge point B2 to each of the output blocks B10. The impact of this modification will be assessed through an ablation study in Section~\ref{ssec:pre-training and ablation study}.

The network produces four output tensors: $\mathbf{F}^{d}_{t_i} \in \mathbb{R}^{N \times M \times C}$, where $i \in \{1,2\}, d \in \{h,v\}$. Each of these tensors contains 1-D filter coefficients for spatially-varying convolution, and $C$ is the filter length. For example, $\mathbf{F}^{h}_{t_i}$ contains horizontal ($h$) filters and $\mathbf{F}^{v}_{t_i}$ contains vertical ($v$) filters for processing $\widetilde{\mathbf{P}}_{t_i}$, $i \in \{1, 2\}$. To apply spatially-varying convolution to pixel $(x,y)$ in $\widetilde{\mathbf{P}}_{t_i}$, we extract (channel-wise) vectors from the corresponding locations in $\mathbf{F}^{h}_{t_i}$ and $\mathbf{F}^{v}_{t_i}$:
\begin{equation}
    \mathbf{f}^{h}_{t_i} = \mathbf{F}^{h}_{t_i}(x,y,:), \qquad
    \mathbf{f}^{v}_{t_i} = \mathbf{F}^{v}_{t_i}(x,y,:),
\label{eq:1-D_filters}
\end{equation}
and then perform the outer product to create a $C \times C$ 2-D filter kernel
\begin{equation}
    \mathbf{K}^{(x,y)}_{t_i} = \mathbf{f}^{v}_{t_i} (\mathbf{f}^{h}_{t_i})^T.
\label{eq:2-D_kernel}
\end{equation}

Finally, pixel value at location $(x,y)$ in color channel $c \in \{1,2,3\}$ in the predicted patch $\widehat{\mathbf{P}}_{t}$ is obtained as
\begin{equation}
\begin{split}
\widehat{\mathbf{P}}_{t}(x,y,c) = & \sum \mathbf{K}^{(x,y)}_{t_1} \circ \widetilde{\mathbf{P}}^{(x,y)}_{t_1}(:,:,c) \ + \\[1ex] & \sum \mathbf{K}^{(x,y)}_{t_2} \circ \widetilde{\mathbf{P}}^{(x,y)}_{t_2}(:,:,c),
\end{split}
\label{eq:proposed_patch}
\end{equation}
where $\circ$ represents the Hadamard product (element-wise multiplication) of the 2-D kernel and a $C \times C$ region of the corresponding input patch centered at $(x,y)$ in color channel $c$. The sum ($\sum$) represents summation of all elements in the Hadamard product. Note that the operation of performing the Hadamard product and summing up the result is called ``local convolution'' and is denoted by symbol $\dot{*}$ in~\cite{Niklaus_ICCV_2017}. We use the same symbol in Fig.~\ref{fig:network_architecture}.
From (\ref{eq:1-D_filters})-(\ref{eq:proposed_patch}) it is easy to see that ``local convolution'' is in fact spatially-varying convolution, because the kernels $\mathbf{K}^{(x,y)}_{t_i}$ depend on $(x,y).$ When performing the Hadamard product in~(\ref{eq:proposed_patch}), samples are taken from the neighborhood of the patch in the corresponding frame, if needed. Near the frame boundary, nearest-neighbor padding is used to fill up the $C \times C$ matrix $\widetilde{\mathbf{P}}^{(x,y)}_{t_i}(:,:,c)$. 

From the video compression point of view, the entire operation is similar to motion compensation from a collocated reference area of size $C \times C$ in the reference frames.  There is no explicit motion model, but adaptive kernels, which are content-dependent, allow pixel values to be moved around to account for motion. Under these conditions, one can think of $\left \lfloor C/2 \right \rfloor$ as being the maximum horizontal/vertical motion component in the system. Importantly, the entire operation (\ref{eq:1-D_filters})-(\ref{eq:proposed_patch}) is differentiable, which allows the network to be trained via gradient descent.

\subsection{Loss function}

The network is trained to predict the original patch $\mathbf{P}_{t}$. To accomplish this, a loss function with several terms is minimized. The first term is the Mean Squared Error (MSE) between the predicted patch $\widehat{\mathbf{P}}_{t}$ and the original patch $\mathbf{P}_{t}$: 
\begin{equation}
\mathcal{L}_{N}= \frac{1}{N \cdot M \cdot 3} \left | \right | \widehat{\mathbf{P}}_{t} - \mathbf{P}_{t}   \left | \right |_{F}^{2}
\label{eq:MSE_loss}
\end{equation}

The second loss term is based on the feature reconstruction loss introduced in~\cite{FeatureReconLoss}.  
According to~\cite{FeatureReconLoss}, this term helps the model improve its prediction of details along the edges, which is important for perceptual quality of the predicted patch. This loss term is defined as
\begin{equation}
\mathcal{L}_{F}=\left | \right | \phi(\widehat{\mathbf{P}}_{t}) - \phi(\mathbf{P}_{t})   \left | \right |_{2}^{2}
\label{eq:feature_loss}
\end{equation}
\noindent where $\phi(\cdot)$ is a feature extraction function. We use the output of the \texttt{relu4\_4} layer of the VGG-19 network~\cite{vgg} as our feature extraction function.

The VGG-19 features used in $\mathcal{L}_{F}$ provide good global features: they are extracted from deep within that network, at the end of the fourth  convolutional block (out of five). However, those features do not capture the local structure of the input signal. For this purpose, we employ another loss term that captures more localized information.  It is based on geometric features, specifically the horizontal and vertical gradients computed as the differences of neighboring pixels. The Mean Absolute Difference (MAD) of the gradients in the predicted and the original patch form the corresponding loss terms: \begin{equation}
\begin{split}
\mathcal{L}_{G_{x}}=\frac{1}{N \cdot M \cdot 3} \sum \left| \frac{\partial}{\partial x}\widehat{\mathbf{P}}_{t} - \frac{\partial}{\partial x}\mathbf{P}_{t} \right| \\[1ex]
\mathcal{L}_{G_{y}}=\frac{1}{N \cdot M \cdot 3} \sum \left| \frac{\partial}{\partial y}\widehat{\mathbf{P}}_{t} - \frac{\partial}{\partial y}\mathbf{P}_{t} \right|
\end{split}
\label{eq:gradient_loss}
\end{equation}
and the summations are carried out over all three color channels. The impact of these loss terms will be examined in the ablation study in  Section~\ref{ssec:pre-training and ablation study}.

Finally, the overall loss is the weighted combination of the loss terms introduced above:
\begin{equation}
\mathcal{L} = \lambda_{N}\cdot\mathcal{L}_{N} + \lambda_{F}\cdot\mathcal{L}_{F} + \lambda_{G}\cdot \left(\mathcal{L}_{G_{x}} + \mathcal{L}_{G_{y}} \right),
\label{eq:final_loss}
\end{equation}
where we empirically selected $\lambda_{N} = \lambda_{F} = 2$ and $\lambda_{G} = 1$ for training. Further details  of training will be discussed in Section~\ref{sec:experimental_results}.

\subsection{Conventional vs. proposed inter prediction}
\label{ssec:texture_prediction}

The proposed DNN is an alternative to conventional inter prediction in that it also uses previously coded frames to predict the currently coded frame.  
Therefore, the proposed approach will compete with conventional prediction in the inter-coded frames in terms of rate distortion (RD) cost. To contrast the two approaches, we first recall the components of the RD cost of conventional inter prediction. 
Suppose a block $\mathbf{B}_{t}$ in the current frame $t$ is to be inter-coded. Depending on whether uni- or bi-directional prediction is used, the motion-compensated prediction $\eta$ has the form
\begin{equation}
\eta (\mathcal{T},\Delta) = 
\begin{cases}
\overline{\mathbf{B}}_{t_1}(x+\Delta x_{t_1}, y+\Delta y_{t_1}), & \text{uni-} \\[1ex]
\begin{aligned}
& \alpha \cdot \overline{\mathbf{B}}_{t_1}(x +\Delta x_{t_1}, y+\Delta y_{t_1}) + \\ 
& (1-\alpha) \cdot \overline{\mathbf{B}}_{t_2}(x+\Delta x_{t_2}, y+\Delta y_{t_2}), 
\end{aligned} 
& \text{bi-}
\end{cases}
\label{eq:conventional_inter}
\end{equation}
where $\overline{\mathbf{B}}_{t_i}$ is the (possibly filtered) motion-compensated prediction from frame $t_i$, $(\Delta x_{t_i}, \Delta y_{t_i})$ is the corresponding motion vector (MV), and $\alpha$ is a weighting parameter, which is set to $0.5$ by default, unless high-level syntax specifies another value~\cite{hevc_std}. $\mathcal{T}$ is the list of time indices of reference frames, which contains only $t_1$ for uni-directional prediction, otherwise both $t_1$ and $t_2$, and $\Delta$ is the list of corresponding MVs, containing one or two MVs, depending on whether prediction is uni- or bi-directional. Note that in bi-directional prediction, the two reference frames are on the opposite sides of the current frame, i.e., $t_1 < t < t_2$. 

The distortion associated with inter prediction is conventionally taken as the Sum of Absolute Differences (SAD) between the current block and its prediction:
\begin{equation}
D_{\textup{Inter}}(\mathcal{T}, \Delta) = \sum \left|\mathbf{B}_{t} - \eta(\mathcal{T}, \Delta) \right|
\label{eq:distortion}
\end{equation}
where the summation is over all pixels in the block. 
Finally, the RD cost of conventional inter prediction is given by
\begin{equation}
J_{\textup{Inter}}(\mathcal{T}, \Delta) = D_{\textup{Inter}}(\mathcal{T}, \Delta) + \lambda \cdot R(\mathcal{T}, \Delta)
\end{equation}
where $R$ denotes a rate function that measures the number of bits required to encode residuals and other necessary parameters, such as MVs, and $\lambda$ is the Lagrange multiplier chosen according to the quantization parameter (QP). RD optimization finds the parameters $(\mathcal{T}, \Delta)$ that lead to the minimum RD cost.

\begin{figure}[t]
    \begin{minipage}[b]{0.48\linewidth}
    \centering
    \includegraphics[width=\textwidth]{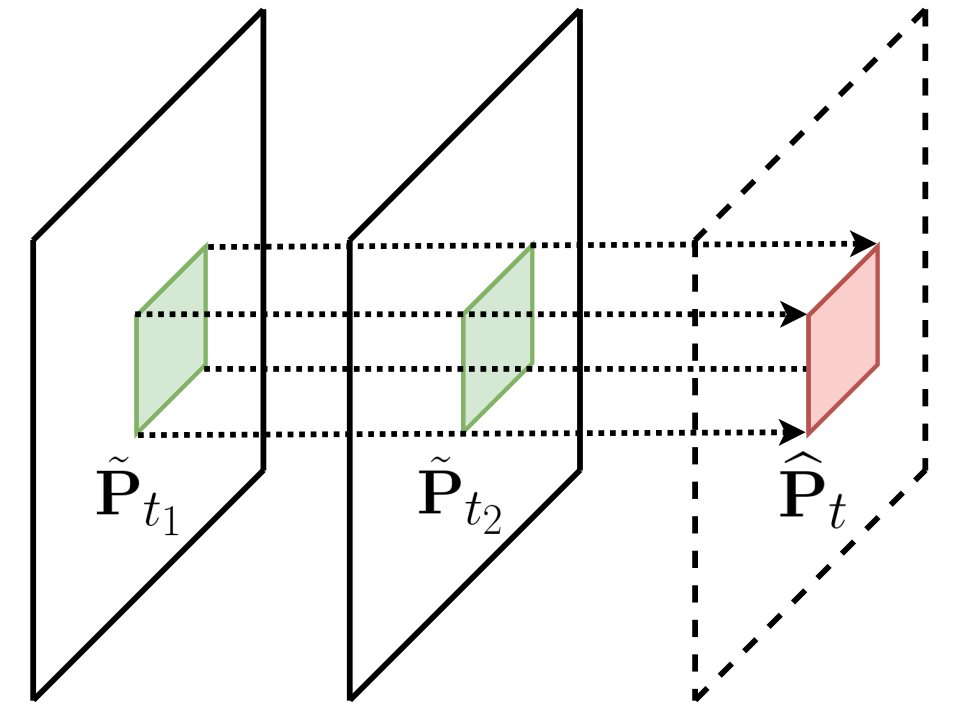}
    \centerline{(a) Uni-directional }\medskip
    \end{minipage}
    \begin{minipage}[b]{0.48\linewidth}
    \centering
    \includegraphics[width=\textwidth]{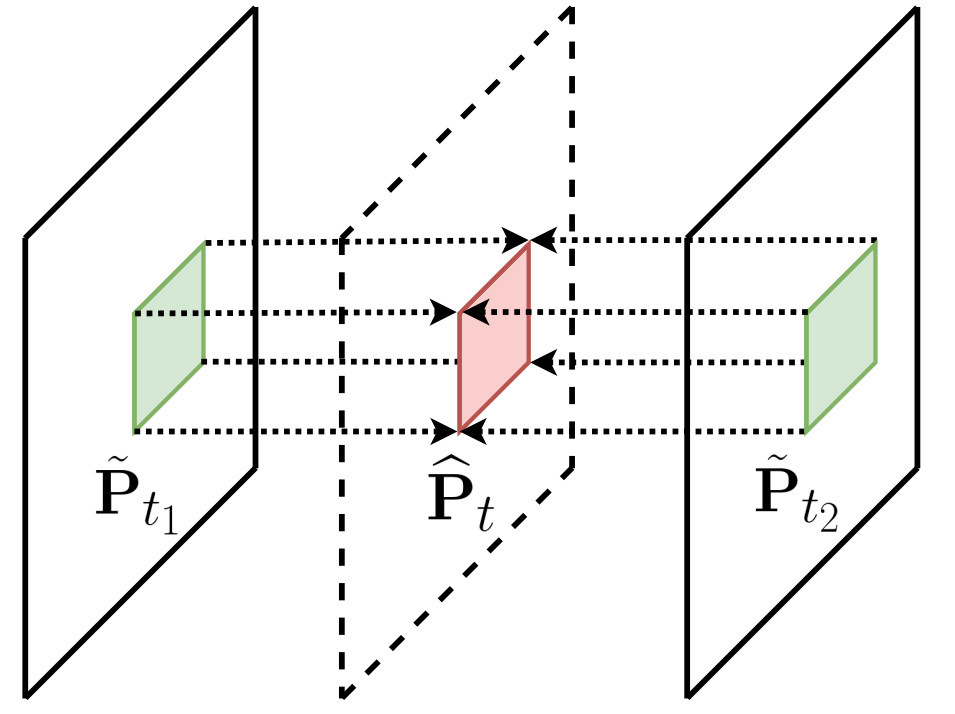}
    \centerline{(b) Bi-directional}\medskip
    \end{minipage}
\caption{Uni- and bi-directional prediction in the proposed method.}
\label{fig:uni-bit-direction}
\end{figure}

On the other hand, the proposed inter prediction operates as shown in Fig.~\ref{fig:uni-bit-direction}. For uni-directional prediction, co-located patches from two preceding frames are taken, and for bi-directional prediction, co-located patches from opposite sides of the current frame are used. Since the patches are co-located, no motion vectors need to be transmitted. Motion information, as well as any possible weighting of contributions from the two reference frames, is effectively encoded in the spatially-varying kernels $\mathbf{K}^{(x,y)}_{t_i}$, which are derived by the DNN from the reference frames, at both encoder and decoder. Apart from the residual, only the flag indicating that the proposed inter prediction is coded. The temporal indices of the reference frames $\mathcal{T} = (t_1, t_2)$ do not need to be transmitted because they can be inferred as follows. In looking for reference frames from DPB, priority is given to bi-directional prediction because, in preliminary testing, we found (unsurprisingly) that bi-directional prediction usually gives better results than uni-directional prediction. Hence, previously coded frames closest to the current frame (up to distance 2) on both sides are selected as references if they are available. If bi-directional prediction is not possible due to unavailability of future coded frames up to distance 2, the two closest previously coded frames preceding the current frame are selected as references.

The distortion associated with the proposed inter prediction is denoted $D_{\textup{DNN}}$ and computed as:
\begin{equation}
D_{\textup{DNN}}(\mathcal{T}) = \sum \left|\mathbf{P}_{t} - \widehat{\mathbf{P}}_{t}\right|
\label{eq:distortion_with_nn}
\end{equation}
where $\widehat{\mathbf{P}}_{t}$ is computed as in~(\ref{eq:proposed_patch}) and the sum is over all pixels in the patch. 
Compared with conventional inter prediction~(\ref{eq:conventional_inter}), the proposed method always exploits two reference patches even for the uni-prediction.  
The RD cost for the proposed method is denoted $J_{\textup{DNN}}$ and is computed as 
\begin{equation}
J_{\textup{DNN}}(\mathcal{T}) = D_{\textup{DNN}}(\mathcal{T}) + \lambda \cdot R(\gamma)
\label{eq:DNN_RD_cost}
\end{equation}
where the rate function measures the number of bits needed to encode the residual and the flag $\gamma$ indicating that the proposed method is used. 

\begin{figure}[t]
    \begin{minipage}[b]{1\linewidth}
    \centering
    \includegraphics[width=\textwidth]{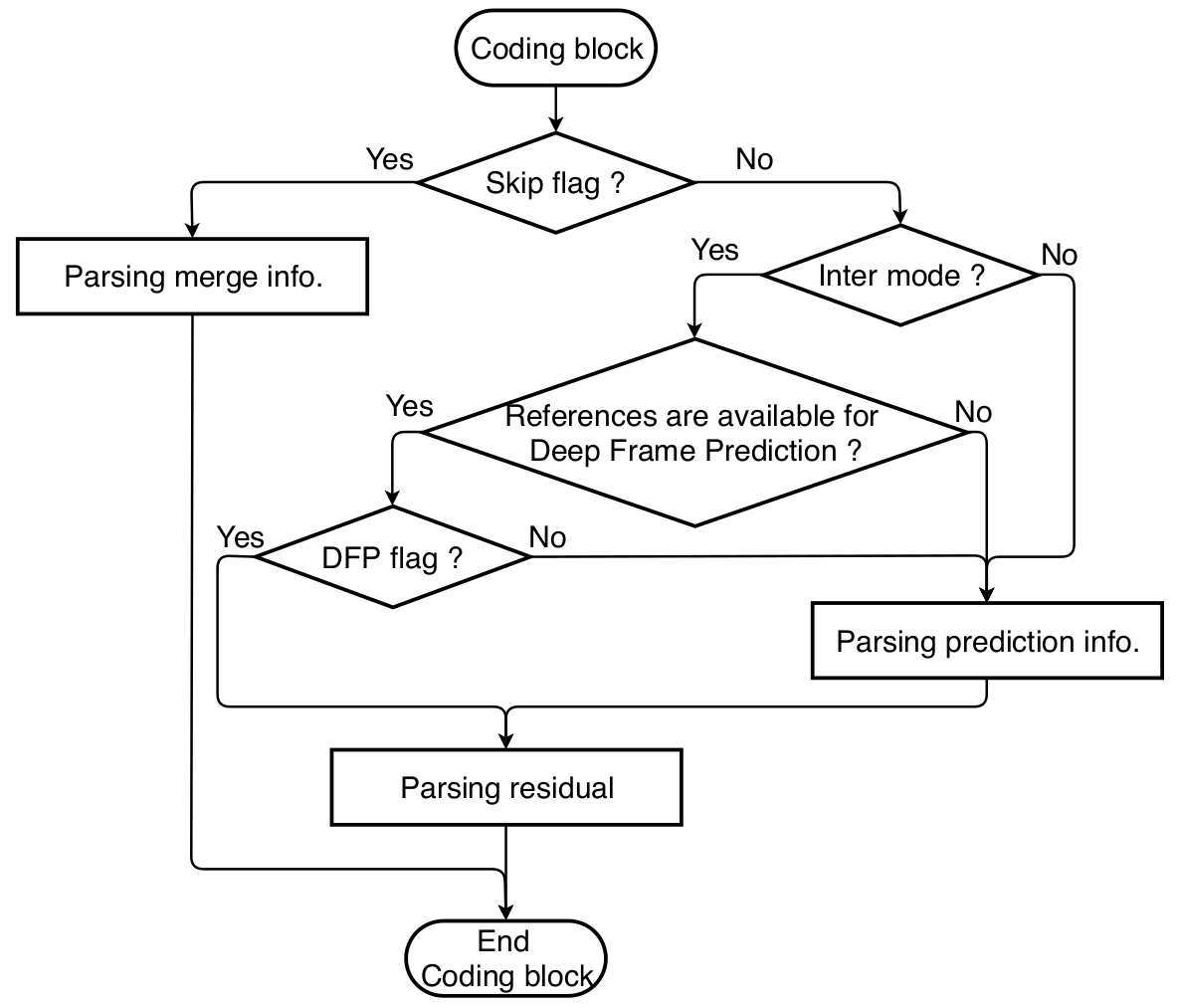}
    \end{minipage}
\caption{Decoding flow with syntax at the coding block level.}
\label{fig:syntax_flow}
\end{figure}

Finally, RD opimization selects the smallest RD cost $J^*$ among the intra, inter, and the proposed DNN-based prediction, as the best coding mode for the current block:
\begin{equation}
J^* = \min(J_{\textup{Intra}}, J_{\textup{Inter}}, J_{\textup{DNN}}).
\label{eq:final_rdcos}
\end{equation}
\noindent In RD optimization, we do not add any bias term to encourage the proposed method to be selected. Instead, we design the syntax carefully so that the proposed method can  straightforwardly compete with other prediction methods in terms of the RD cost based on the distortion and the actual bits to be coded. Fig.~\ref{fig:syntax_flow} shows the simplified decoding flow with syntax at the coding block level, where the DFP flag stands for deep frame prediction (i.e., the proposed method). Considering that reference frames for the proposed method might not be available in certain cases in the hierarchical-B structure~\cite{hevc_sze}, the DFP flag is parsed only if these references are available. In addition, in order to minimize overhead bits for the proposed method, parsing the DFP flag is done after parsing intra/inter mode. Therefore, the DFP flag is not used in intra blocks. This is also the case for the inter block where the references for the proposed method are unavailable. When the DFP flag is set, only the residual bits are parsed.

Note that the blocks coded by the proposed method do not have any MVs associated with them. While this saves bits compared to conventional inter coding, it may inadvertently increase the bits used for other, conventionally inter-coded blocks in the same frame, because their MVs may have fewer neighboring MVs from which MV prediction can be made. In order to mitigate this, we examined using motion estimation (at both encoder and decoder, without sending any extra information) for DNN-predicted blocks in order to assign MVs to them for the purpose of MV prediction of other, conventionally coded MVs. However, this did not lead to any rate saving, which leads us to conclude that MVs estimated for DNN-predicted blocks were rarely, if ever, selected as MV predictors. Hence, we abandoned this idea, and the final codec did not incorporate this feature. 

\section{DNN training}
\label{sec:network_training}
In this section, we detail the process of our DNN training. Since the performance of the trained network directly affects the coding gain of the proposed frame prediction, choosing the effective learning strategy is very important. Some of the training choices, such as the optimizer and learning rate, are based on the exploration by Niklaus \textit{et al.}~\cite{Niklaus_ICCV_2017}. Accordingly, we employed AdaMax~\cite{kingma2014adam} with default parameter values, and started training from scratch with the learning rate of 0.001. We also set the mini-batch size to 16. However, there are also some key differences relative to~\cite{Niklaus_ICCV_2017}. Since the focus application of our DNN is video coding, we used YUV sequences, rather than RGB. Also, our training samples are collected from~\cite{xiph_org}, which  contains diverse raw sequences at various frame rates. We found raw video sequences with 25, 30, 50, and 60 frames per second (fps) and various resolutions from 352$\times$240 (SIF) to 1920$\times$1080 (FullHD). Considering the diversity of resolutions and frame rates in our data, we chose a two-part training strategy consisting of pre-training and fine tuning. 

\subsection{Pre-training with ablation study}
\label{ssec:pre-training and ablation study}
The purpose of pre-training is to find, on a relatively small dataset, reasonably good network structure and its weights from which large-scale fine tuning can start.  
For pre-training, we employ a total of 27 sequences: 25 for training and another 2 for validation. Resolutions of these 27 sequences were either 352$\times$240 (SIF) or 352$\times$288 (CIF). In order to train a model that is able to operate at various QP values, 
we adopted compression augmentation~\cite{dfc_for_collab_object_detection}.  
The key strategy in compression augmentation is to introduce the quantization error into the training process by compressing and decompressing data with multiple quantization parameters. Therefore, the network learns to be resilient to quantization errors of different magnitudes, arising from different quantization parameters. The following describes the compression augmentation process for our pre-training.
First, the chosen sequences were encoded at various QPs: 20, 22, 24, ..., 44, for a total of 13 QP values. For each QP value, three different coding configurations were used: Low delay (LD), Low delay P (LP), and Random access (RA). During training, either a raw sequence or one of the three coding configurations was chosen randomly, and if the choice was a coded sequence, then one QP was selected randomly from the 13 QP values available. 

A training sample consists of three collocated patches of size 128$\times$128 randomly chosen from a triplet of frames within a given sequence. One epoch covers a total of 6,158 training frame triplets and 698 validation triplets. In addition to compression augmentation, which was done offline, we used several forms of online augmentation, namely horizontal and vertical flipping, and reversing the frame order, so that the network gets exposed to a variety of motions. The network was trained for 500 epochs on NVIDIA GeForce GTX 1080 GPU with 11 GB RAM.

\begin{figure}[t]
    \begin{minipage}[b]{1\linewidth}
    \centering
    \includegraphics[width=\textwidth]{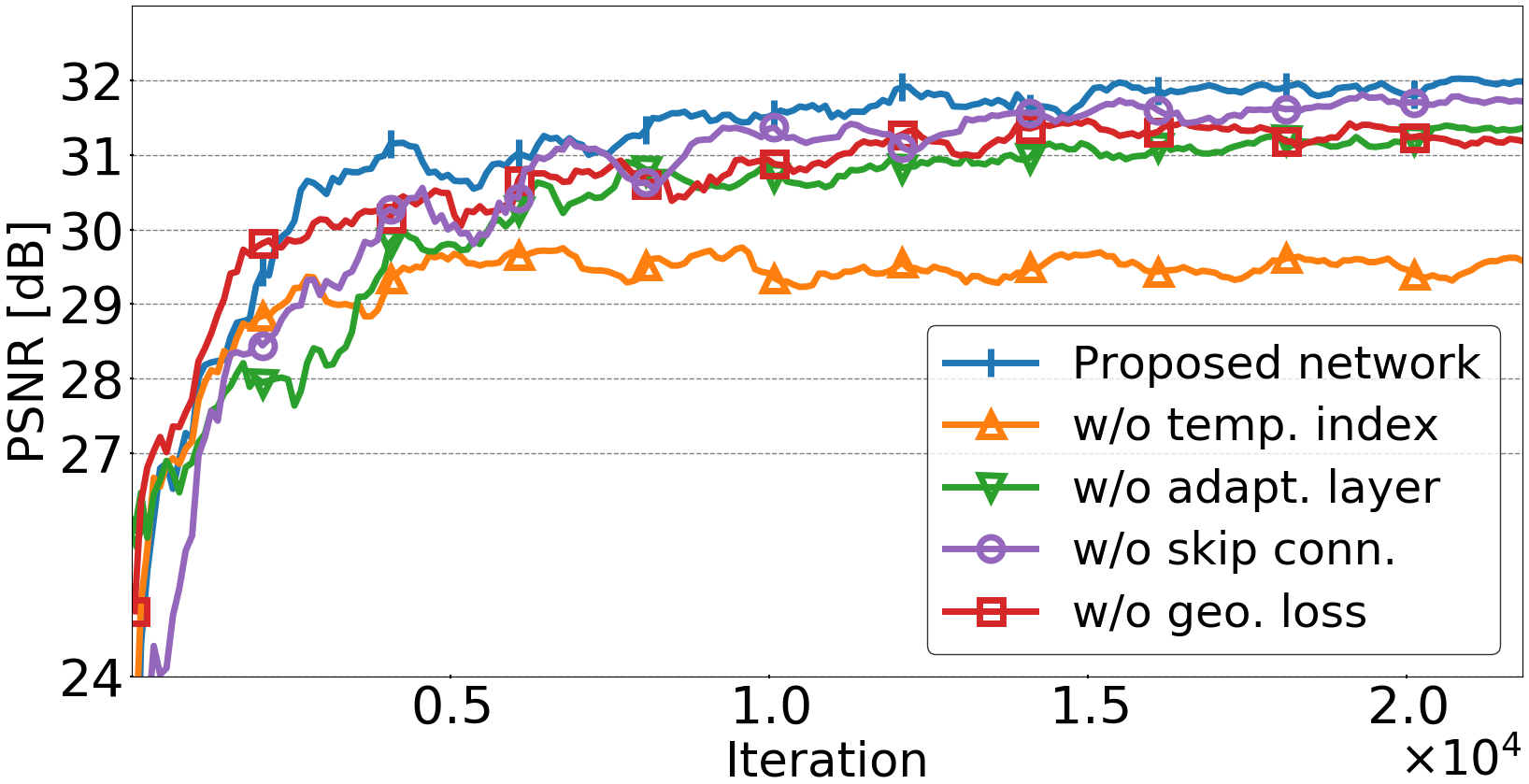}
    \end{minipage}
\caption{Validation PSNR for the proposed network and its ablated versions.} 
\label{fig:temporal_channel_effective}
\end{figure}

In order to assess the the contribution of individual components of the proposed network, we carry out the ablation study at the pre-training stage. For this purpose, we remove individual components of the network, one at a time, and examine the performance of the resulting ablated network during pre-training, by monitoring the Peak Signal to Noise Ratio (PSNR) of predicted patches in the validation set. The components whose impact is examined in this study are: temporal index channel, adaptation layer B1, skip connections from B2 to B10, and geometric loss~(\ref{eq:gradient_loss}). 
The results are shown in Fig.~\ref{fig:temporal_channel_effective} as PSNR vs. mini-batch iteration.
The blue curve in Fig.~\ref{fig:temporal_channel_effective} is the performance of the full (non-ablated) proposed network.

In order to examine the effect of the temporal index channel $\mathbf{T}_{t_i}$, we set it to a dummy constant, same for both input patches. The resulting performance is shown as the orange curve in Fig.~\ref{fig:temporal_channel_effective}. As seen, without the correct temporal index, the performance of the network drops by about 2 dB in PSNR, which is the highest drop in this ablation study. Hence, we conclude that the temporal index channel is the most important component among those examined in this study. Second, we remove B1 blocks from the proposed network  such that the two input patches are stacked and directly fed to B2. The results are shown as a green curve in Fig.~\ref{fig:temporal_channel_effective}. According to the results, prediction performance degrades by approximately 1 dB compared with the complete network. Third, we remove skip connections from the merge point B2 to each of the outputs B10, and the corresponding results are shown as the purple curve in Fig.~\ref{fig:temporal_channel_effective}. According to the results, excluding these skip connections reduces prediction performance by about 0.5 dB compared with the complete network. Hence, these skip connections offer useful support to temporally-variant processing by allowing outputs to depend on the temporal index of the reference frames. Lastly, we examined the efficacy of the geometric loss terms~(\ref{eq:gradient_loss}). In this case, the network architecture is complete, as shown in Fig.~\ref{fig:network_architecture}, but it is trained with $\lambda_{G}=0$  in~(\ref{eq:final_loss}). The results are shown as the red curve in Fig.~\ref{fig:temporal_channel_effective}. As seen, the prediction performance degrades by about 1 dB in this case, hence local geometric features seem to be quite useful in achieving good prediction.

\subsection{Fine tuning}
Fine tuning starts from the network weights obtained in pre-training. In this stage, a lager dataset and a more sophisticated training strategy was used. 
First, patch triplets were drawn randomly from the training sequences with resolutions ranging from SIF to FullHD. Patches of size 150$\times$150 were initially drawn, and from these patches, collocated windows of 128$\times$128 were extracted. Motion augmentation~\cite{Niklaus_CVPR_2017} in the form of shifting the reference windows was used to increase the diversity of motions seen by the network.  
 
To further enrich the diversity of  training samples, we eliminated ``monotonous'' triplets -- those whose  entropy in any patch was less than 3.5 bits per pixel, and those that did not exhibit any pixel value change between any pair of patches. We also estimated optical flow between the (temporally) first and last patch in the triplet using the Coarse2Fine method~\cite{pathakCVPR17learning} and eliminated patch triplets with zero variance in optical flow magnitudes.  
We also excluded triplets with extreme motion, where the largest optical flow magnitude was larger than the patch diagonal.  
Finally, the training set was formed with 27,360 triplets, and the validation set had 3,040 triplets. Training was done for 500 epochs starting from the weights obtained in pre-training, and using the same data augmentation strategy as in pre-training.

In addition to the DNN that supports both uni-and bi-directional prediction, we trained two other DNNs with the same architecture: one specifically for uni-directional prediction, the other specifically for bi-directional prediction. The same data and the same strategy -- pre-training followed by fine tuning -- was used for these two DNNs as well. These two DNNs will be used in the next section for comparison purposes.

\section{Experimental results}
\label{sec:experimental_results}

The setup used for experimental evaluation of the proposed approach is shown in Table~\ref{tbl:exp_env}. Two groups of experiments were carried out: evaluation of frame prediction performance in Section~\ref{sec:prediction_exp} and evaluation of video coding performance in Section~\ref{sec:coding_exp}. Frame prediction experiments examine only the quality of predicted frames, whereas video coding experiments  take into account compressibility of prediction residuals and any side information needed by the decoder to reconstruct the frames.

\begin{table}[t]
\centering
\caption{Experimental environment}
\label{tbl:exp_env}
\begin{tabular}{@{}ll@{}}
\toprule
Item             & Specification                            \\ \midrule
Processor        & Intel(R) Core(TM) i7-7800X CPU @ 3.50GHz \\
GPU              & GeForce GTX 1080 with 11 GB RAM          \\
Operating system & Ubuntu-16.04                             \\
HEVC version     & HM-16.20                                 \\
DNN framework         & Pytorch-gpu-0.4.1-py36 with CUDA 9.0     \\ \bottomrule
\end{tabular}
\end{table}

\subsection{Frame prediction performance}
\label{sec:prediction_exp}

\begin{table}[t]
\centering
\caption{Average Y-PSNR of predicted frames}
\label{tbl:avg_y_psnr_comparison}
\smallskip\noindent
\resizebox{\linewidth}{!}{%

\setlength{\tabcolsep}{1.7pt}
\begin{tabular}{@{}cccccc|cccc@{}}
\toprule
\multirow{2}{*}{Sequence} & \multirow{2}{*}{fps} & \multicolumn{4}{c}{Uni-directional (dB)} & \multicolumn{4}{c}{Bi-directional (dB)} \\ \cmidrule(l){3-10} 
                      &   & HEVC  & ~\cite{Niklaus_ICCV_2017}  & Sep.  & Comb.  & HEVC  & ~\cite{Niklaus_ICCV_2017}  & Sep.  & Comb.  \\ \midrule
BQSquare              & 60  & \textcolor{red}{\textit{36.05}} & 22.62  & \textbf{34.45}  & 33.12   & \textcolor{red}{\textit{37.40}} & \textbf{37.19} &         36.97   & 36.78  \\
BasketballPass        & 50  & \textcolor{red}{\textit{37.44}}  & 21.15  & 30.28  & \textbf{30.33}  & \textcolor{red}{\textit{38.76}} &         31.57  & \textbf{33.21}  & 32.99  \\
RaceHorses            & 30  & \textcolor{red}{\textit{35.79}}  & 18.96  & \textbf{27.38}  & 27.27  & \textcolor{red}{\textit{37.26}} &         29.35  & \textbf{30.15}  & 30.04  \\
ParkScene             & 24  & \textcolor{red}{\textit{40.87}}  & 26.55  & \textbf{35.96}  & 35.83  & \textcolor{red}{\textit{41.51}} &         36.74  & \textbf{38.51}  & 38.30  \\ \bottomrule

\end{tabular}}
\end{table}

\begin{figure*}[t]
    \begin{minipage}[b]{0.197\linewidth}
    \centering
    \includegraphics[width=\textwidth]{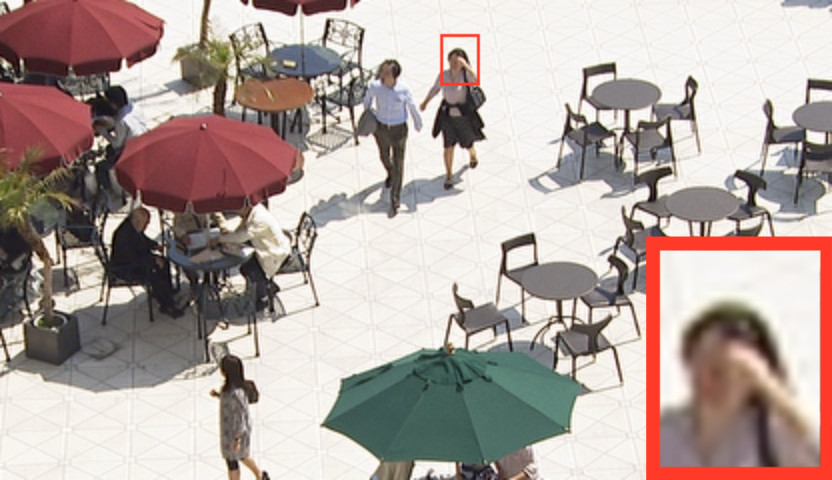}
    \end{minipage}
    \hspace{-0.14cm}
    \begin{minipage}[b]{0.197\linewidth}
    \centering
    \includegraphics[width=\textwidth]{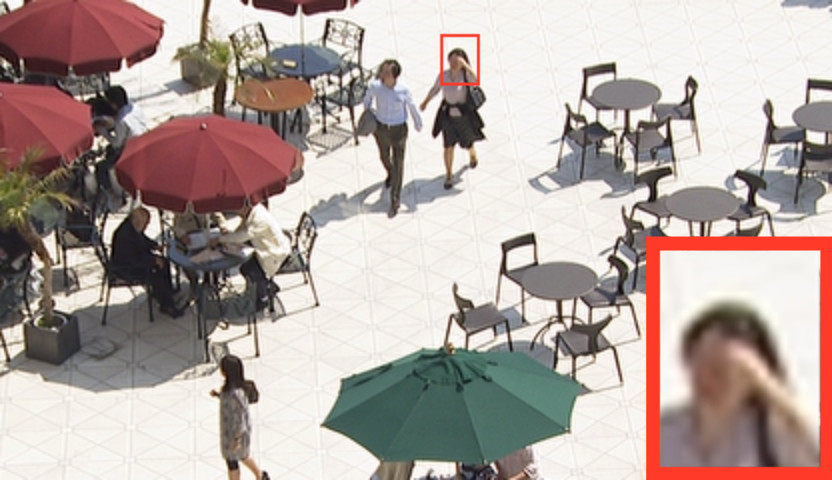}
    \end{minipage}
    \hspace{-0.14cm}
    \begin{minipage}[b]{0.197\linewidth}
    \centering
    \includegraphics[width=\textwidth]{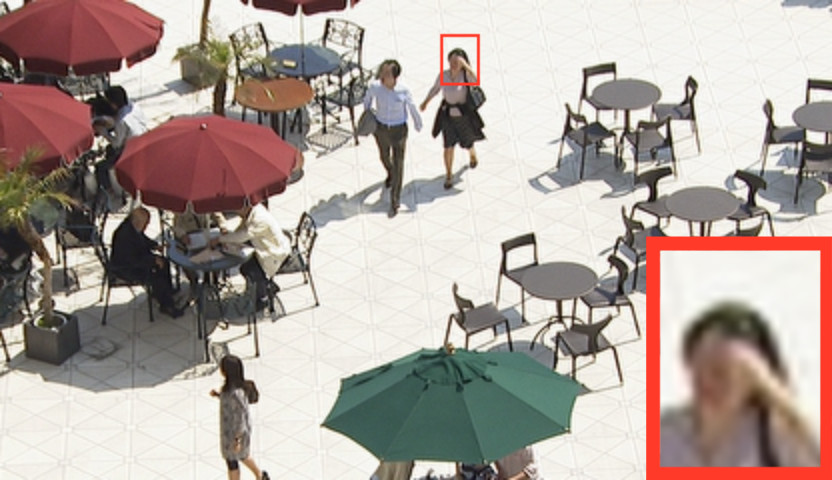}
    \end{minipage}
    \hspace{-0.14cm}
    \begin{minipage}[b]{0.197\linewidth}
    \centering
    \includegraphics[width=\textwidth]{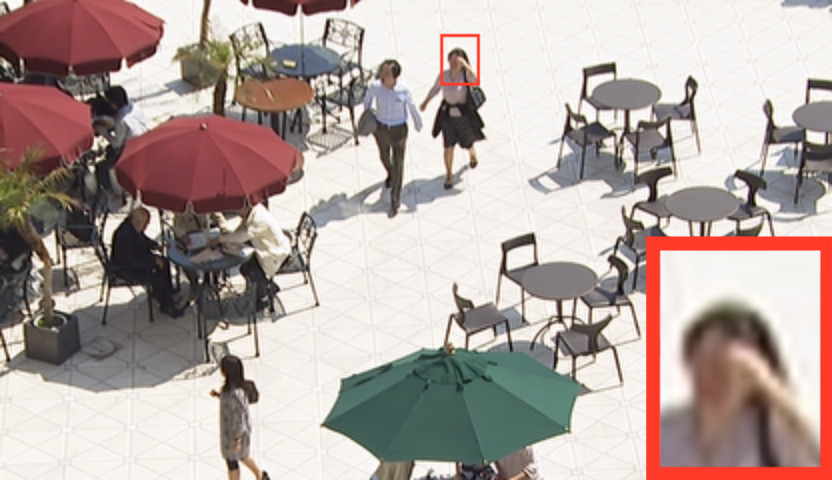}
    \end{minipage}
    \hspace{-0.14cm}
    \begin{minipage}[b]{0.197\linewidth}
    \centering
    \includegraphics[width=\textwidth]{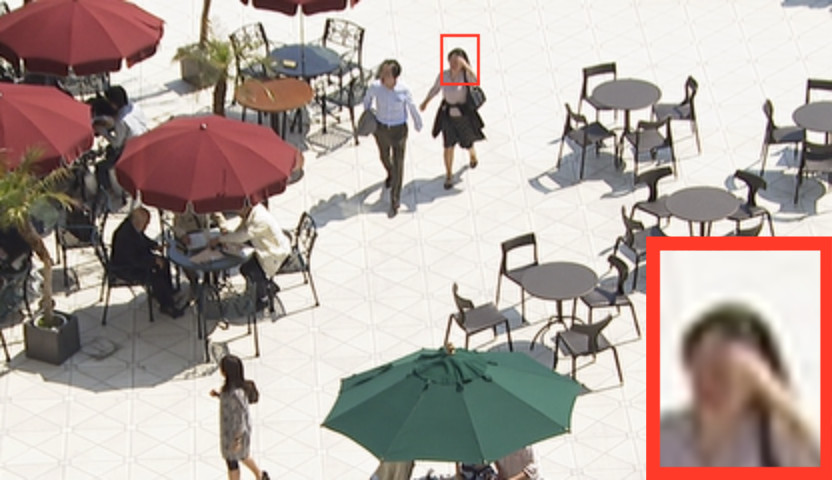}
    \end{minipage}
    \hspace{-0.2cm}
    \hfill
    
    \vspace{0.08cm}
    \begin{minipage}[b]{0.197\linewidth}
    \centering
    \includegraphics[width=\textwidth]{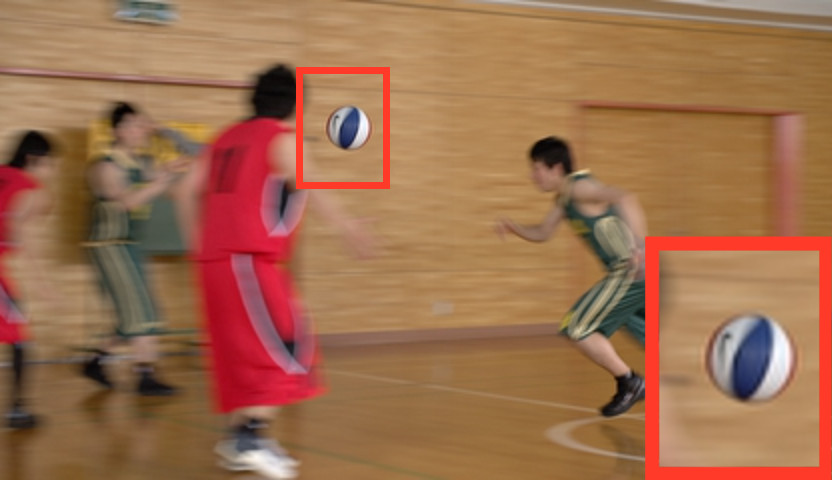}
    \end{minipage}
    \hspace{-0.14cm}
    \begin{minipage}[b]{0.197\linewidth}
    \centering
    \includegraphics[width=\textwidth]{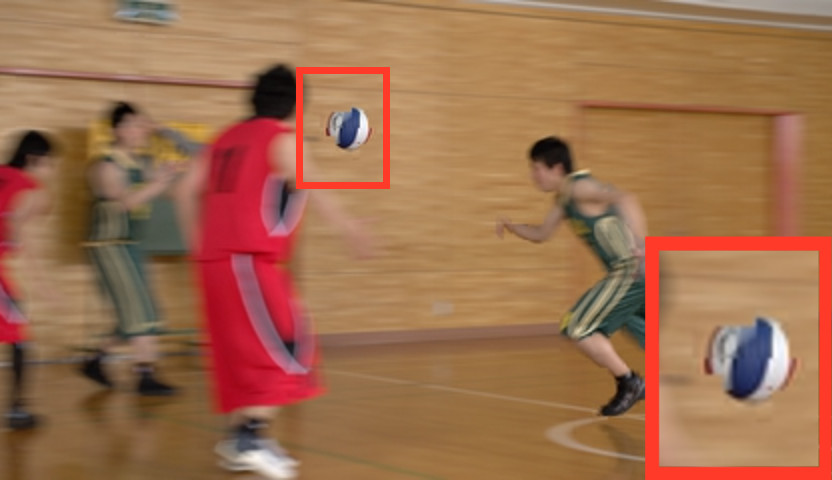}
    \end{minipage}
    \hspace{-0.14cm}
    \begin{minipage}[b]{0.197\linewidth}
    \centering
    \includegraphics[width=\textwidth]{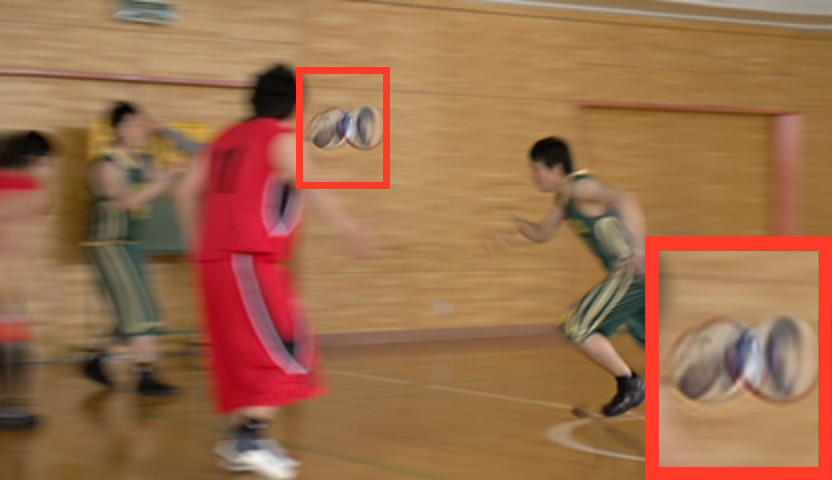}
    \end{minipage}
    \hspace{-0.14cm}
    \begin{minipage}[b]{0.197\linewidth}
    \centering
    \includegraphics[width=\textwidth]{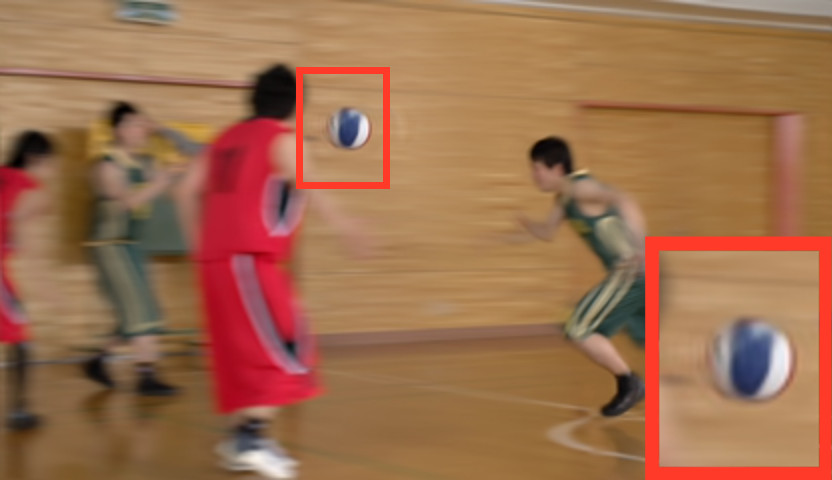}
    \end{minipage}
    \hspace{-0.14cm}
    \begin{minipage}[b]{0.197\linewidth}
    \centering
    \includegraphics[width=\textwidth]{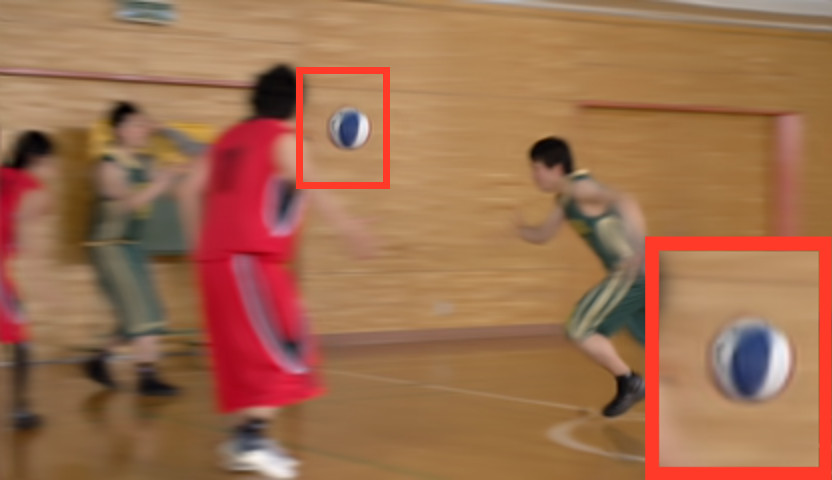}
    \end{minipage}
    \hspace{-0.2cm}
    \hfill
    
    \vspace{0.08cm}
    \begin{minipage}[b]{0.197\linewidth}
    \centering
    \includegraphics[width=\textwidth]{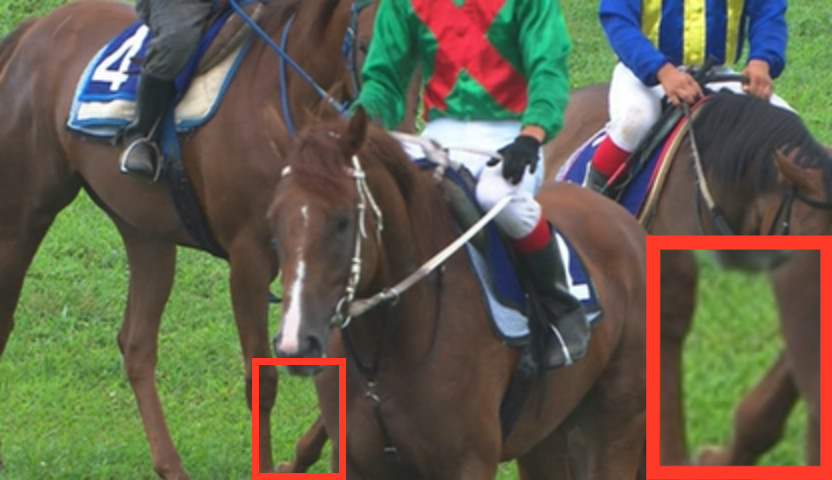}
    \end{minipage}
    \hspace{-0.14cm}
    \begin{minipage}[b]{0.197\linewidth}
    \centering
    \includegraphics[width=\textwidth]{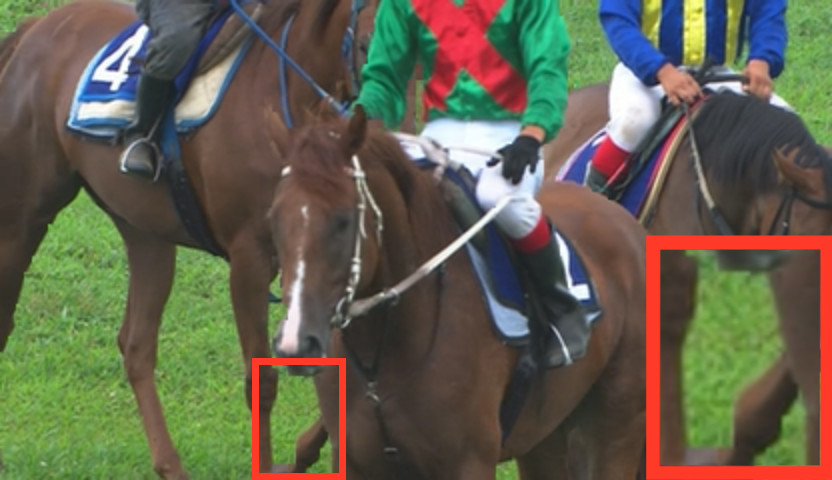}
    \end{minipage}
    \hspace{-0.14cm}
    \begin{minipage}[b]{0.197\linewidth}
    \centering
    \includegraphics[width=\textwidth]{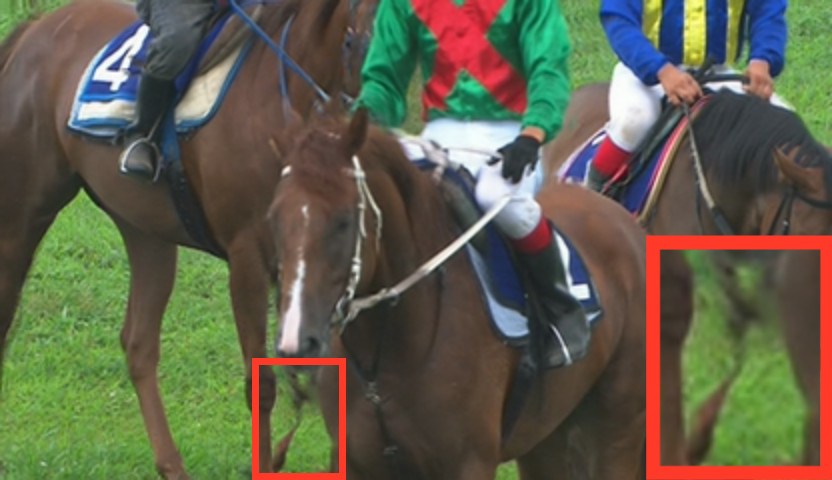}
    \end{minipage}
    \hspace{-0.14cm}
    \begin{minipage}[b]{0.197\linewidth}
    \centering
    \includegraphics[width=\textwidth]{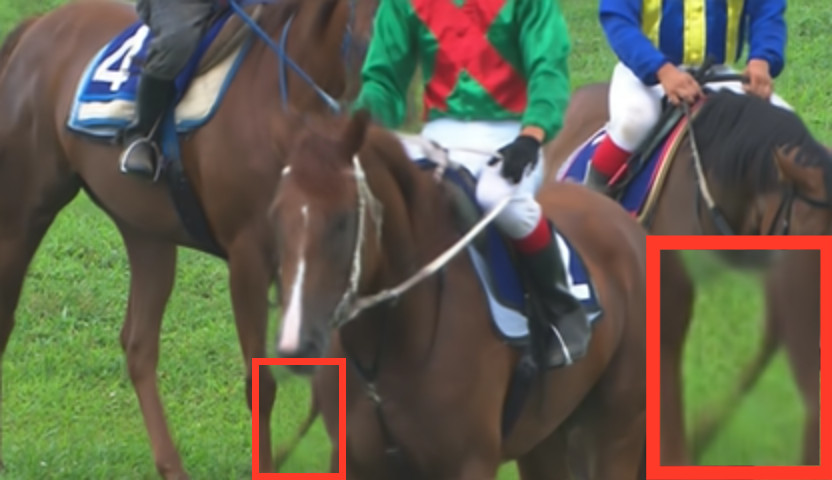}
    \end{minipage}
    \hspace{-0.14cm}
    \begin{minipage}[b]{0.197\linewidth}
    \centering
    \includegraphics[width=\textwidth]{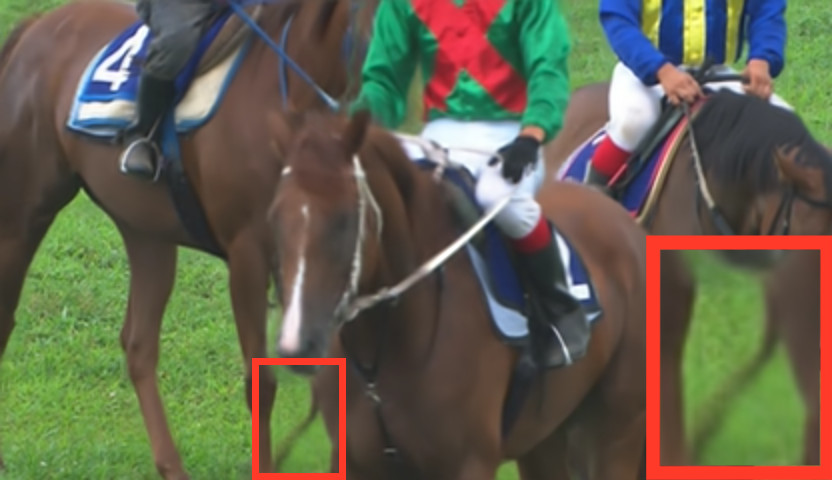}
    \end{minipage}
    \hspace{-0.2cm}
    \hfill
    
    \vspace{0.08cm}
    \begin{minipage}[b]{0.197\linewidth}
    \centering
    \includegraphics[width=\textwidth]{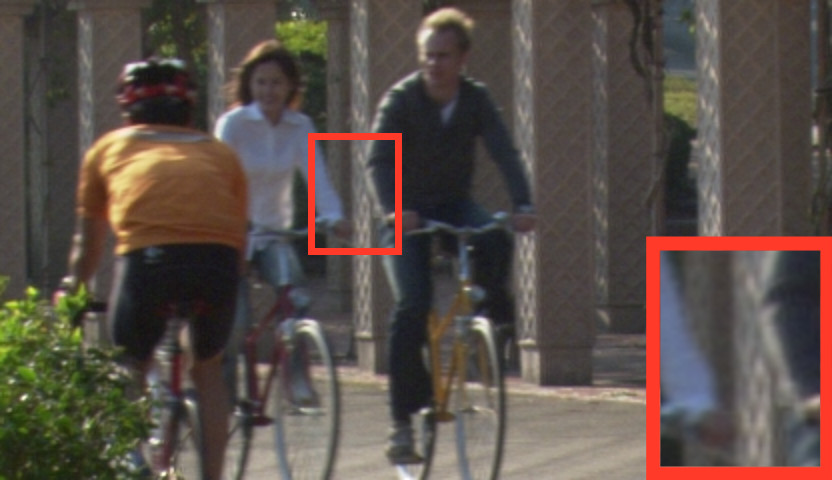}
    \centerline{(a)}
    \end{minipage}   
    \hspace{-0.14cm}
    \begin{minipage}[b]{0.197\linewidth}
    \centering
    \includegraphics[width=\textwidth]{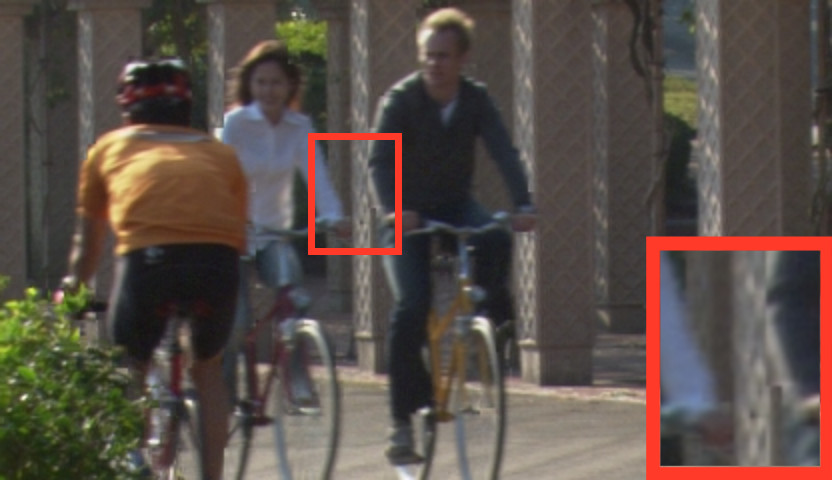}
    \centerline{(b)}
    \end{minipage}
    \hspace{-0.14cm}
    \begin{minipage}[b]{0.197\linewidth}
    \centering
    \includegraphics[width=\textwidth]{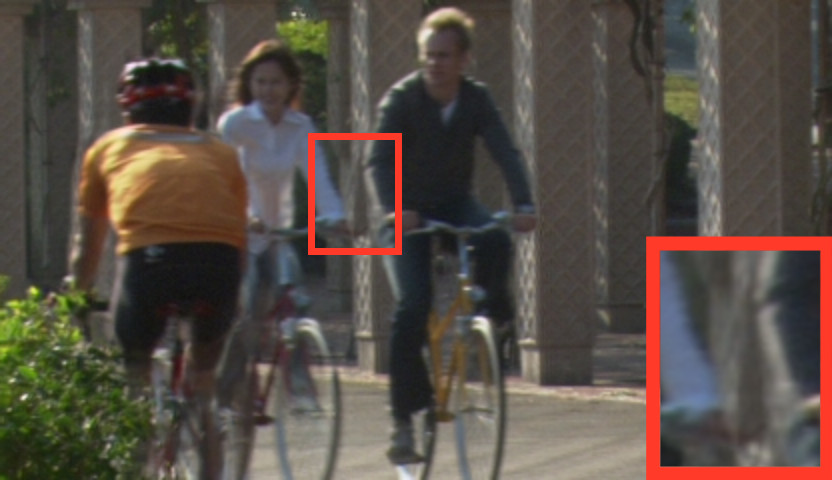}
    \centerline{(c)}
    \end{minipage}
    \hspace{-0.14cm}
    \begin{minipage}[b]{0.197\linewidth}
    \centering
    \includegraphics[width=\textwidth]{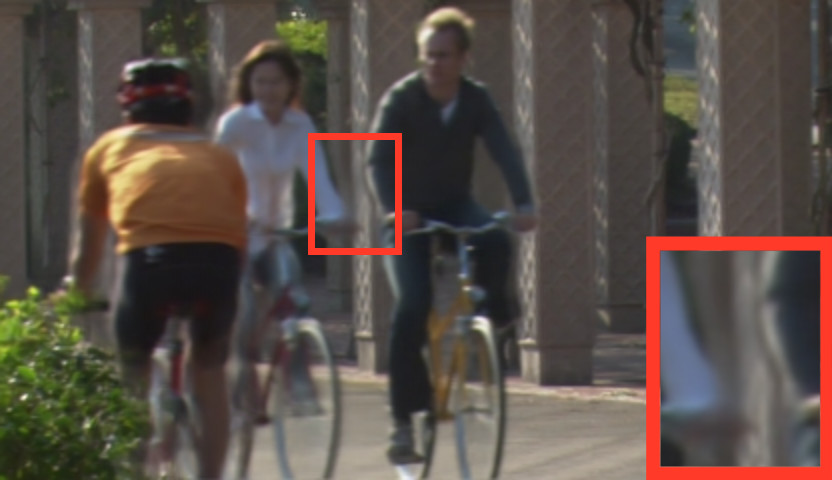}
    \centerline{(d)}
    \end{minipage}
    \hspace{-0.14cm}
    \begin{minipage}[b]{0.197\linewidth}
    \centering
    \includegraphics[width=\textwidth]{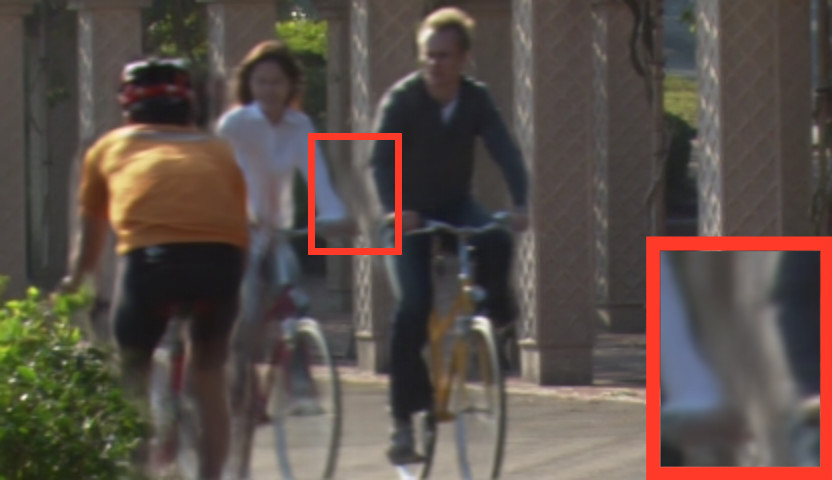}
    \centerline{(e)}
    \end{minipage} 
    \hspace{-0.2cm}
    \hfill
    
\vspace{-0.1cm}
\caption{Visual comparison of bi-directionally predicted frames: (a) original, (b) frame predicted by HEVC Inter-frame coding, (c) frame predicted by~\cite{Niklaus_ICCV_2017}, (d) frame predicted by our separately trained bi-directional DNN, (e) frame predicted by our combined uni-/bi-directional DNN.}
\label{fig:bi_direction_visual_comparison}
\end{figure*}

\begin{figure}[t]
    \begin{minipage}[b]{0.47\linewidth}
    \centering
    \includegraphics[width=\textwidth]{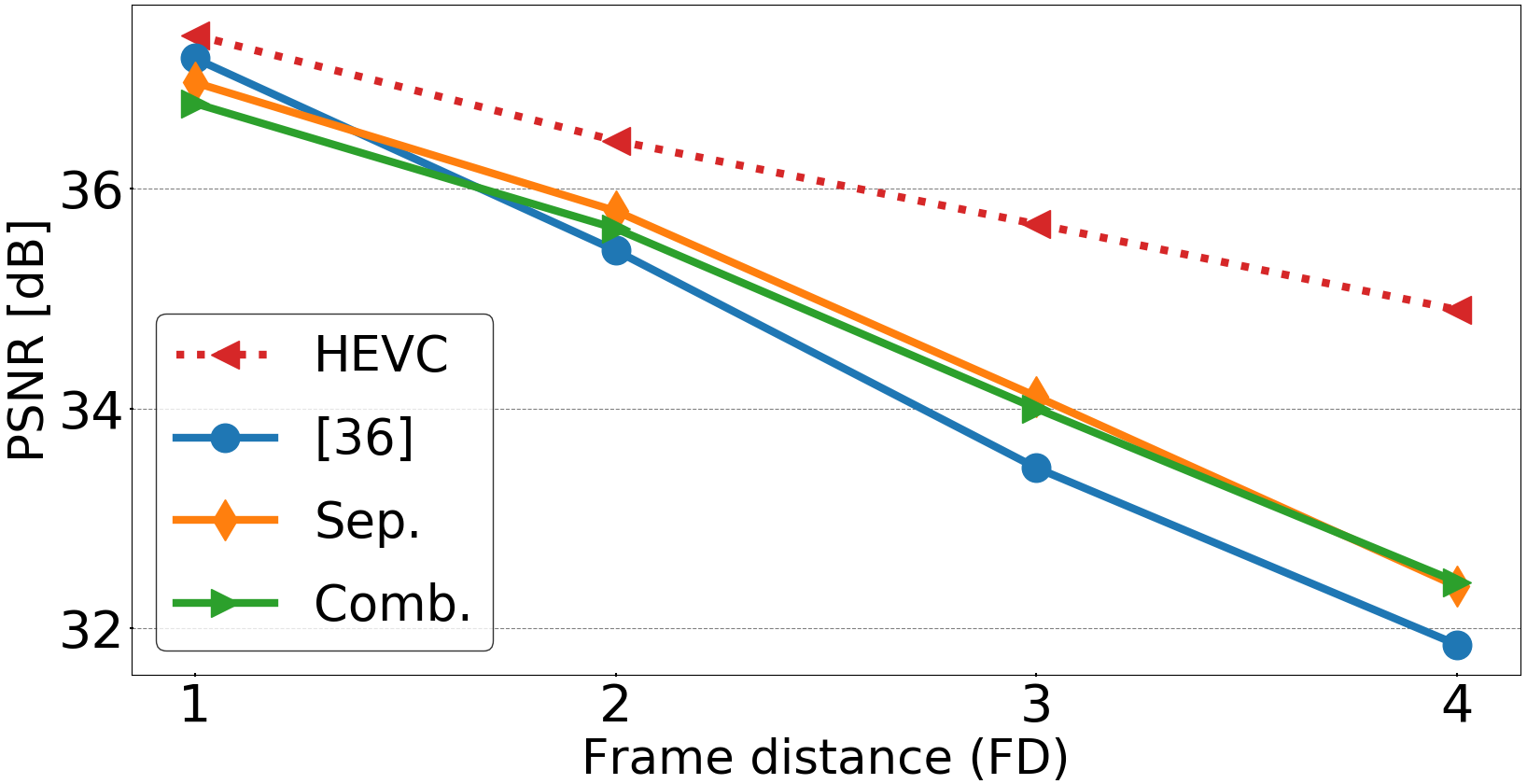}
    \centerline{BQSquare}
    \end{minipage}
    \hspace{-0.1cm}
    \begin{minipage}[b]{0.47\linewidth}
    \centering
    \includegraphics[width=\textwidth]{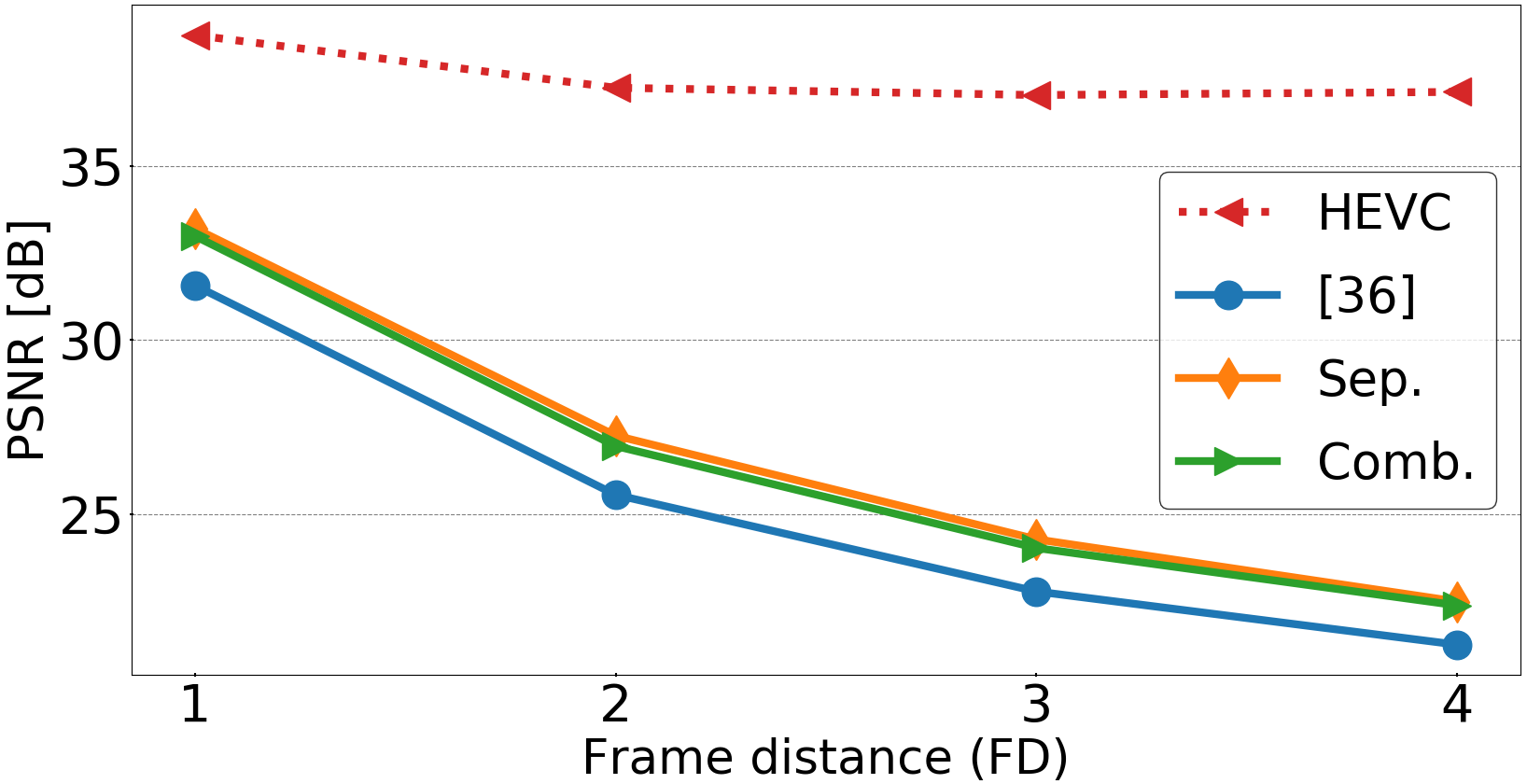}
    \centerline{BasketballPass}
    \end{minipage}
    \vspace{0.5cm}
    
    \begin{minipage}[b]{0.47\linewidth}
    \centering
    \includegraphics[width=\textwidth]{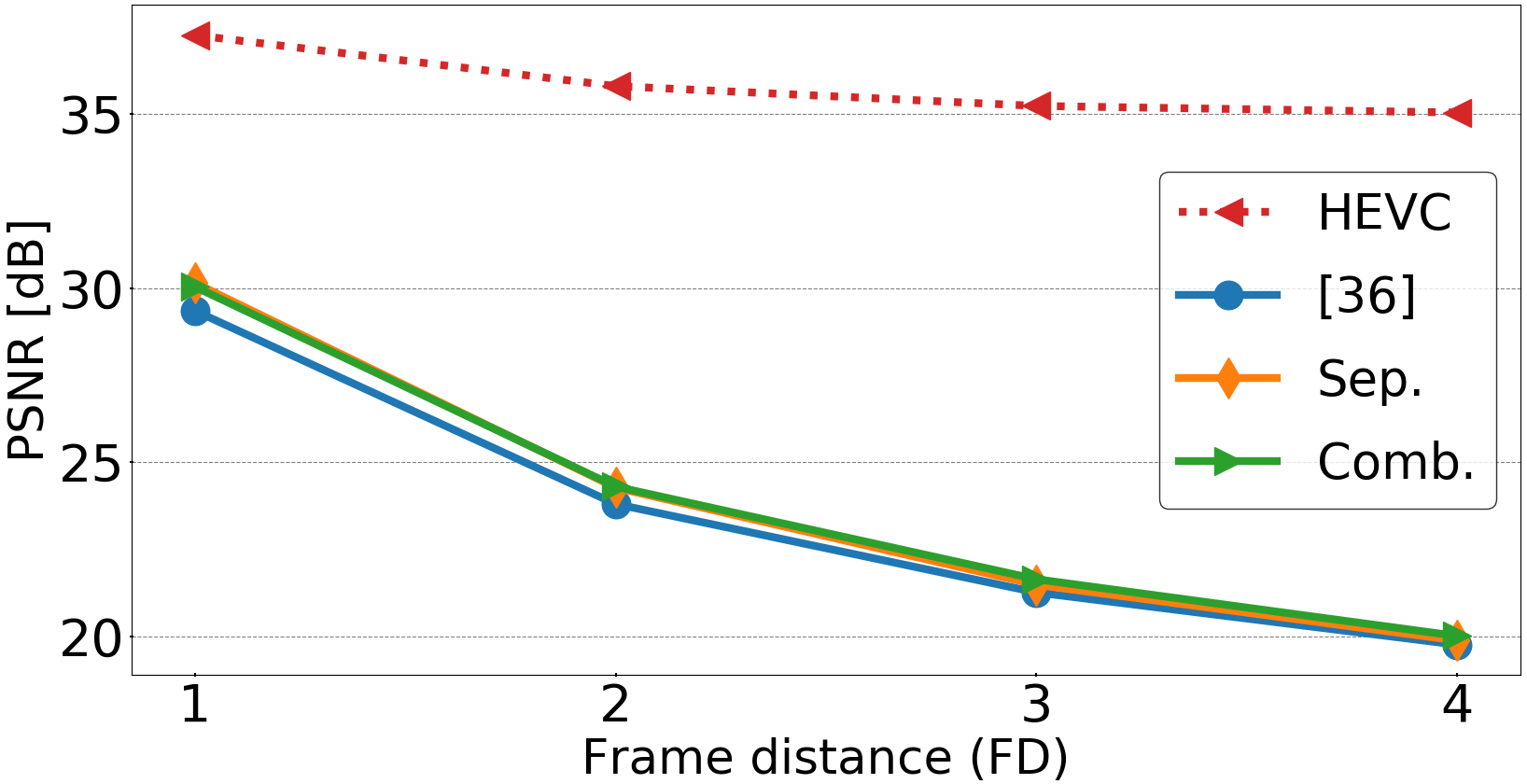}
    \centerline{RaceHorses}
    \end{minipage}
    \hspace{-0.1cm}
    \begin{minipage}[b]{0.47\linewidth}
    \centering
    \includegraphics[width=\textwidth]{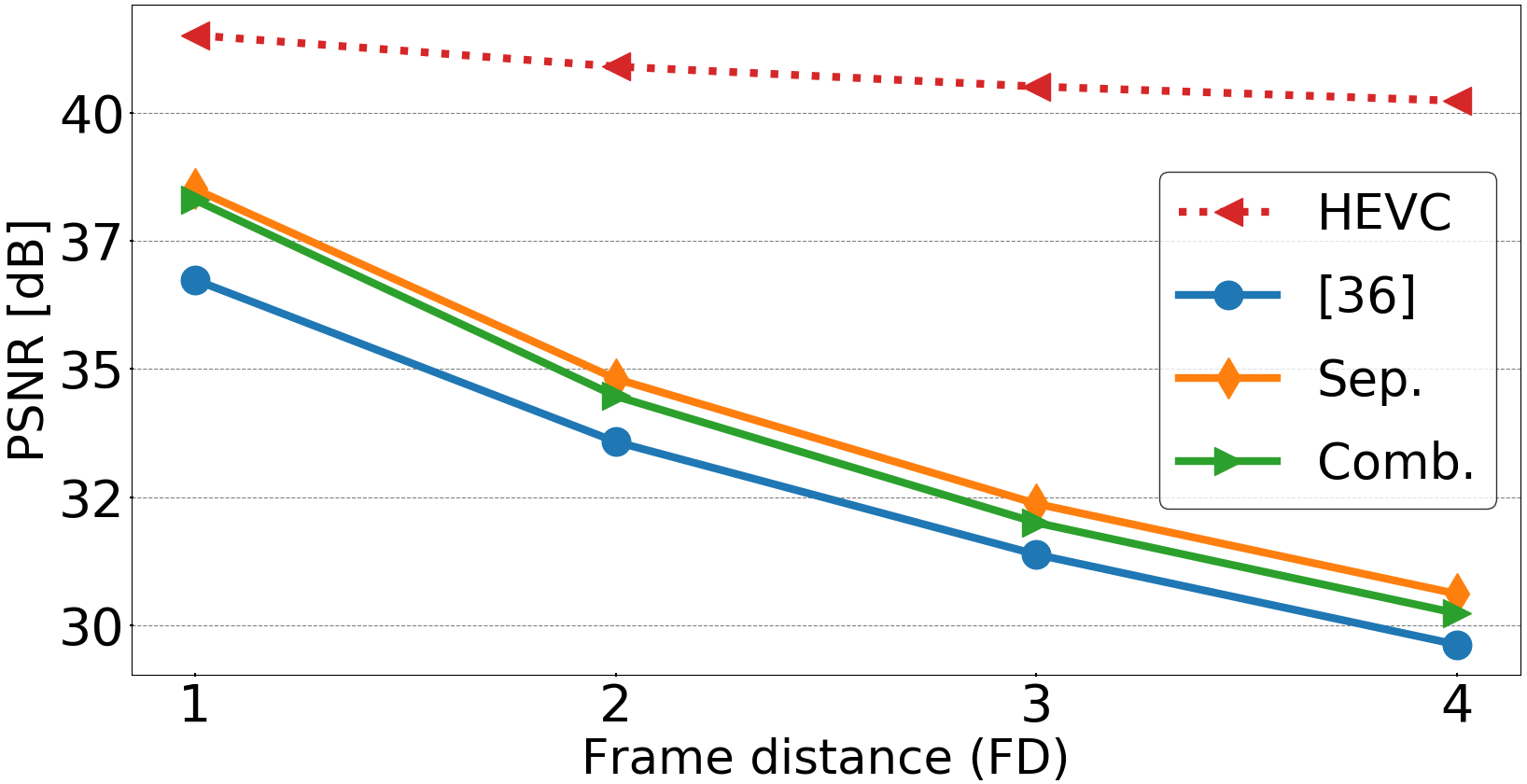}
    \centerline{ParkScene}
    \end{minipage}
    \vspace{-0.1cm}
\caption{Bi-directional prediction with varying frame distance.}
\label{fig:bi_directional_comparison}
\end{figure}
 
We compare the frame prediction performance of our three networks -- one for combined uni-/bi-directional prediction and two for separate uni- and bi-directional prediction -- with the reference network~\cite{Niklaus_ICCV_2017}, which is used in~\cite{hevc_with_sep_conv_for_ra}, as well as HEVC Lossless~\cite{hevc_std}, which allows which allows using original frames as references. The comparison is carried out on the four HEVC test sequences~\cite{hevc_ctc} listed in Table~\ref{tbl:avg_y_psnr_comparison}. 
Note that the resolution of all sequences is 416$\times$240, except for ParkScene, which is 1920$\times$1080 (FullHD). This sequence was cropped to 416$\times$240 starting with the position (600, 600) in the FullHD frame as the top left corner. For uni-directional prediction, the temporal indices of reference frames were $(t_1,t_2)=(t-2, t-1)$ and for the bi-directional case they were $(t_1, t_2) = (t-1, t+1)$. 

Table~\ref{tbl:avg_y_psnr_comparison} presents the average PSNR of the Y-component of predicted frames for each model and HEVC. Our DNN that performs combined uni-/bi-directional prediction is denoted ``Comb.'' in the table, while the DNNs for separate uni- and bi-directional prediction are denoted ``Sep.'' As seen in the table, the frames predicted by HEVC show higher PSNR than any of the DNNs (ours and~\cite{Niklaus_ICCV_2017}), because the predicted samples use motion vectors as side information to minimize the prediction error. However, motion vectors need to be transferred to the decoder, so they become additional cost in RD optimization. According to~\cite{stankowski2015analysis}, the bits needed to represent motion information usually account for about 10-30\% of the total bits spent on an inter frame. Eventually, HEVC pays a heavy price for motion-compensated prediction. By contrast, our method does not require any motion information at the decoder so, in the end, it achieves better RD performance, as demonstrated in the next section. 

The uni-directional performance of~\cite{Niklaus_ICCV_2017} is quite low, understandably, because the network was not trained for this case. But our models outperform~\cite{Niklaus_ICCV_2017} in bi-directional prediction in three out of four sequences. The only exception is BQSquare, with very low motion, where all models do fairly well. Among our models, while separately trained models are clearly better than the combined model, the difference is not large, less than 0.5 dB. Most importantly, the combined model outperforms~\cite{Niklaus_ICCV_2017} in most bi-directional cases, even though it can support both uni- and bi-directional prediction, whereas~\cite{Niklaus_ICCV_2017} supports only bi-directional prediction.  

Several visual examples of bi-directional prediction are shown in Fig.~\ref{fig:bi_direction_visual_comparison}. The BQSquare sequence in the first row is characterized by camera panning at a distance, with high frame rate. Thus the difference between consecutive frames is very small, and the the predicted frame is fairly accurate (around 37 dB) for all models. Only subtle differences can be found in the zoomed-in face region shown in the bottom right of each frame.  
In the second row, the BasketballPass sequence has much more complicated motion, and the differences in predicted frames are easily noticeable. The basketball is poorly predicted by HEVC and especially the network from~\cite{Niklaus_ICCV_2017}, as shown in the second and third column, respectively. Even though HEVC achieves high average PSNR on this sequence, distortion on high-speed objects such as the basketball is visible. Meanwhile, our models last two columns seem to do a better job on the basketball.  
With the RaceHorses sequence in the third row, all network models struggle to reconstruct the horse's leg, but our models made better effort to preserve context around the leg.  
Finally, with the ParkScene sequence in the last row, the proposed models again provides better prediction than~\cite{Niklaus_ICCV_2017}, which suffers from the noticeable warp on the pillar along the the arm. The frame predicted by HEVC is good overall, despite some small blocking artifacts. 

For bi-directional prediction to be used in  hierarchical B coding in the RA configuration, prediction is sometimes required from more distant frames, not just the immediately preceding and succeeding frame. Therefore, in Fig.~\ref{fig:bi_directional_comparison}, we evaluate bi-directional prediction accuracy from a previous and a future frame at various (symmetric) distances from the current frame. The figure shows average Y-PSNR (in dB) of the predicted frame vs. frame distance, averaged over all frames in the corresponding sequences where such bi-directional prediction is possible. As above, HEVC prediction provides the highest PSNR with the benefit of motion vectors. The proposed DNNs, both the separately trained bi-directional DNN and the combined uni-/bi-directional DNN, outperform the network from~\cite{Niklaus_ICCV_2017} in all cases except BQSquare at frame distance of 1. At the same time, except in BQSquare at frame distance of 1, our separately trained bi-directional DNN is slightly better than our combined uni-/bi-directional DNN, most notably in ParkScene at larger frame distances.

\subsection{Video coding performance}
\label{sec:coding_exp}

\begin{table*}[t]
\centering
\caption{BD-Bitrate relative to HM-16.20 over three common test conditions}
\label{tbl:overall_coding_performance}
\smallskip\noindent
\resizebox{\linewidth}{!}{%
\setlength\tabcolsep{3.5pt}
\begin{tabular}{@{}cccccc|ccc|ccc|ccc|ccc|ccc@{}}
\toprule
\multirow{3}{*}{Class} & \multirow{3}{*}{Sequence} & \multirow{3}{*}{fps} & \multicolumn{6}{c}{Low delay P (LP)}                          & \multicolumn{6}{c}{Low delay (LD)}                            & \multicolumn{6}{c}{Random access (RA)}                        \\ \cmidrule(l){4-21} 
                       &                            &                      & \multicolumn{3}{c}{Sep. (\%)} & \multicolumn{3}{c}{Comb. (\%)} & \multicolumn{3}{c}{Sep. (\%)} & \multicolumn{3}{c}{Comb. (\%)} & \multicolumn{3}{c}{Sep. (\%)} & \multicolumn{3}{c}{Comb. (\%)} \\ \cmidrule(l){4-21} 
                       &                            &                      & Y      & U      & V      & Y      & U      & V     & Y      & U      & V      & Y      & U      & V     & Y      & U      & V      & Y      & U      & V      \\ \midrule
\multirow{2}{*}{A}     & PeopleOnStreet             & \multirow{2}{*}{30}  & --5.8  & --4.0  & --4.4  & --5.2  & --3.5  & --4.8 & --4.4  & --2.2  & --2.7  & --3.7  & --2.1  & --3.3 & --4.0  & --5.7  & --5.6  & --4.2  & --5.7  & --5.9 \\
                       & Traffic                    &                      & --3.6  & --3.8  & --3.1  & --3.5  & --4.0  & --2.9 & --2.5  & --2.4  & --2.3  & --2.1  & --2.7  & --2.5 & --2.5  & --2.7  & --2.3  & --2.2  & --2.5  & --2.0 \\ \midrule
\multirow{5}{*}{B}     & BQTerrace                  & 60                   & --4.2  & --3.6  & --1.4  & --4.0  & --2.2  &   0.1 & --1.3  & --2.0  & --1.4  & --1.1  & --1.2  &   0.7 & --2.0  & --0.5  & --0.4  & --1.6  &   0.0  &   0.0 \\
                       & BasketballDrive            & \multirow{2}{*}{50}  & --2.8  & --5.6  & --3.7  & --2.8  & --5.1  & --3.5 & --1.1  & --2.6  & --1.8  & --0.9  & --2.0  & --1.3 & --1.2  & --2.0  & --1.5  & --1.0  & --1.7  & --1.2 \\
                       & Cactus                     &                      & --6.9  &--10.3  & --6.0  & --6.0  & --9.1  & --5.5 & --3.7  & --7.1  & --4.7  & --2.8  & --5.7  & --3.4 & --2.8  & --5.2  & --3.4  & --3.0  & --4.9  & --3.3 \\
                       & Kimono                     & \multirow{2}{*}{24}  & --5.1  & --7.3  & --3.7  & --5.7  & --7.8  & --4.3 & --2.1  & --4.4  & --2.5  & --1.8  & --3.9  & --2.5 & --0.9  & --1.6  & --1.0  & --0.9  & --1.6  & --0.9 \\
                       & ParkScene                  &                      & --3.3  & --4.4  & --2.8  & --3.2  & --5.0  & --2.8 & --2.1  & --3.0  & --2.5  & --1.9  & --3.4  & --2.2 & --1.9  & --2.4  & --1.8  & --1.6  & --2.0  & --1.6 \\ \midrule
\multirow{4}{*}{C}     & BQMall                     & 60                   & --6.0  & --7.3  & --6.7  & --5.3  & --7.1  & --6.3 & --4.5  & --6.3  & --5.7  & --3.6  & --5.8  & --4.6 & --4.2  & --4.9  & --4.7  & --3.5  & --4.2  & --3.9 \\
                       & BasketballDrill            & \multirow{2}{*}{50}  & --2.9  & --5.2  & --3.0  & --2.9  & --5.2  & --3.4 & --1.6  & --3.4  & --1.6  & --1.4  & --3.1  & --1.8 & --1.5  & --2.8  & --2.6  & --1.5  & --2.7  & --2.4 \\
                       & PartyScene                 &                      & --3.3  & --3.9  & --3.8  & --2.9  & --3.1  & --3.1 & --2.0  & --3.0  & --3.1  & --1.6  & --2.3  & --2.3 & --4.5  & --4.0  & --3.8  & --3.7  & --3.8  & --3.4 \\
                       & RaceHorsesC                & 30                   & --1.1  & --1.5  & --1.9  & --1.0  & --1.5  & --1.9 & --0.8  & --1.1  & --1.5  & --0.6  & --0.8  & --1.2 & --0.7  & --1.2  & --1.5  & --0.6  & --1.0  & --1.2 \\ \midrule
\multirow{4}{*}{D}     & BQSquare                   & 60                   & --3.2  & --0.6  &   1.5  & --1.4  & --2.7  & --0.3 & --1.6  & --0.6  & --3.0  & --0.8  & --2.1  & --1.8 & --3.3  & --0.7  & --0.4  & --2.3  & --0.1  &   0.0 \\
                       & BasketballPass             & \multirow{2}{*}{50}  & --4.4  & --5.7  & --4.6  & --4.2  & --5.7  & --4.2 & --3.2  & --4.3  & --3.6  & --2.9  & --3.7  & --3.1 & --4.1  & --5.6  & --4.7  & --3.5  & --4.7  & --3.4 \\
                       & BlowingBubbles             &                      & --4.1  & --5.2  & --5.3  & --3.4  & --5.0  & --4.9 & --2.7  & --4.5  & --5.4  & --2.2  & --3.8  & --4.4 & --4.3  & --4.0  & --3.8  & --3.7  & --3.8  & --3.4 \\
                       & RaceHorses                 & 30                   & --1.8  & --2.7  & --3.2  & --1.6  & --2.4  & --2.5 & --1.3  & --2.3  & --2.6  & --1.1  & --2.1  & --1.9 & --1.6  & --2.8  & --3.0  & --1.5  & --2.4  & --2.5 \\ \midrule
\multirow{3}{*}{E}     & FourPeople                 & \multirow{3}{*}{60}  & --10.1 &--11.9  &--11.6  & --9.8  &--11.3  &--10.5 & --7.0  & --9.3  &--10.1  & --6.5  & --8.5  & --8.7 & \multicolumn{3}{c}{--}       & \multicolumn{3}{c}{--} \\
                       & Johnny                     &                      & --8.6  & --9.1  & --7.5  & --7.8  & --8.8  & --8.3 & --4.9  & --7.2  & --7.4  & --4.2  & --6.4  & --5.5 & \multicolumn{3}{c}{--}       & \multicolumn{3}{c}{--} \\
                       & KristenAndSara             &                      & --8.7  &--22.2  & --9.1  & --7.9  &--10.6  & --9.7 & --5.4  & --8.5  & --7.4  & --4.6  & --7.4  & --5.9 & \multicolumn{3}{c}{--}       & \multicolumn{3}{c}{--} \\ \midrule
\multicolumn{3}{c}{Average}                                                & --4.8  & --5.7  & --4.5  & --4.4  & --5.6  & --4.4 & --2.9  & --4.1  & --3.8  & --2.4  & --3.7  & --3.1 & --2.6  & --3.1  & --2.7  &  --2.3 & --2.7  & --2.3        \\ \midrule
\multicolumn{3}{c}{$\Delta T_{Enc}$}         & \multicolumn{3}{c}{163\%}    & \multicolumn{3}{c}{164\%} & \multicolumn{3}{c}{146\%}    & \multicolumn{3}{c}{145\%}    & \multicolumn{3}{c}{150\%}    & \multicolumn{3}{c}{149\%}    \\ \midrule
\multicolumn{3}{c}{$\Delta T_{Dec}$}         & \multicolumn{3}{c}{16,563\%} & \multicolumn{3}{c}{16,562\%} & \multicolumn{3}{c}{15,415\%} & \multicolumn{3}{c}{15,410\%} & \multicolumn{3}{c}{11,389\%} & \multicolumn{3}{c}{11,488\%} \\ \bottomrule
\end{tabular}}
\end{table*}

In this section, we evaluate the performance of the proposed frame prediction 
in video coding. The benchmark is the latest video coding standard, High Efficiency Video Coding (HEVC)~\cite{hevc_std}. The proposed method competes with HEVC inter and intra prediction in inter frames in terms of the RD cost~(\ref{eq:final_rdcos}), where the proposed method is selected when $J_{\textup{DNN}}<J_{\textup{Inter}}, J_{\textup{Intra}}$. As shown in Table~\ref{tbl:exp_env}, the proposed prediction method is implemented based on HM-16.20 and the DNN is implemented in Pytorch. Python Embedding Library\footnote{https://docs.python.org/3/extending/} is used to embed the DNN into HM-16.20. During video coding, forward operation of the DNN to perform frame prediction is executed on the GPU (Table~\ref{tbl:exp_env}). However, due to the memory limitations of the GPU, it is impossible to generate the prediction frame larger than 832$\times$480 ($\approx$WVGA) resolution. Therefore, larger frames are split into multiple tiles with a maximum size of WVGA. Then tile-wise prediction is performed and the predicted frame is assembled from predicted tiles.

For evaluation, three coding configurations (LD, LP, and RA) that allow inter-frame coding under the common test conditions (CTC) with the common test sequences (CTS)~\cite{hevc_ctc} are employed with four quantization parameters QP $\in \{22, 27, 32, 37\}$. In the LD and LP configurations, only uni-directional prediction is available using two previously coded frames before updating the reference picture set (RPS)~\cite{hevc_mathias} of the current frame. 
For the RA configuration,  bi-directional prediction is performed when the coded frames with a maximum difference of $\pm$2 from the POC of the current frame are available before updating the RPS. Larger distances can be allowed, but the computational complexity increases while prediction performance reduces at larger distances, so we felt the POC difference of up to 2 is a good compromise.

Table~\ref{tbl:overall_coding_performance} shows the coding performance compared to the HEVC for various configurations. We measure the coding performance using  BD-Bitrate~\cite{Bjontegaard}. We show the performance of both the separate uni- and bi-directional models (``Sep.'' in the table), as well as the combined uni-/bi-directional model (``Comb.'' in the table). 
The proposed method reduces the bitrate for the luminance component in all test cases. There is some increase for the chrominance components of one sequence (BQTerrace,  LP and LD configurations for Comb),  but considering that the chrominance components are much smaller than luminance, this increase is negligible compared to the overall savings. And on average, the bitrate of chrominance components across all sequences is reduced. For the LP configuration, the proposed method achieves the largest bits savings of up to 10.1\% and 9.8\% with the separate model and the combined model, respectively, for the sequence FourPeople. In general, significant bit reductions are shown in larger resolution sequences in Classes A, B and E. 

Compared with the LP configuration, the coding gains are somewhat lower in the LD configuration, especially in Class B sequences. However, gains in Class E (teleconference) sequences are still strong. Similarly, in the RA configuration, the coding gains are somewhat less than in LP, but comparable to the coding gains in the LD configuration. 
The largest luminance coding gains in this case are 4.5\% with Sep and 4.2\% with Comb, achieved on the sequences PartyScene and PeopleOnStreet, respectively. 
Moreover, significant bit savings were shown in Class C and D. 

Overall, the proposed method achieves higher coding gain in the LP configuration compared to LD and RA configurations. This is likely due to the fact that it uses two reference frames and content-adaptive filters derived by the DNN, compared to HEVC, which uses a single reference frame in this configuration. The gain also depends on the sequence complexity, as will be discussed later in this section.  
As could be expected, the combined uni-/bi-directional DNN (Comb) offers somewhat lower gains than separate uni- and bi-directional DNNs (Sep) by about 0.3\%-0.5\%, but it supports both forms of prediction, so it allows a simpler video coding system design, which may be preferred in many applications.

The last two rows of the table show the difference in encoding and decoding time,  defined as 
\begin{equation}
\Delta T_p = \frac{T_{Proposed, p}}{T_{HM, p}} \times 100\%
\label{eq:coding_complexity}
\end{equation}
where $p\in\{Enc, Dec\}$, and $T_{Proposed,p}$ and $T_{HM,p}$ represent the the total elapsed time for the proposed method and the original HM-16.20, respectively. Due to the complexity of running a deep model, the encoding run time increases by 45-65\%, depending on the configuration. However, the decoding time increases much more significantly. This is because decoding time is small to start with, and the complexity of frame prediction with the proposed method is symmetric (i.e., it takes the same amount of time) at the encoder and the decoder, so compared to an initially small decoding time, running a deep model adds significant complexity. As with other envisioned applications of deep models, run-time complexity is one of the bottlenecks, and clever solutions will need to be found to get this technology into end-user products.

\begin{table}[t]
\centering
\caption{Average BD-Bitrate with MS-SSIM relative to HM-16.20 over three common test condition}
\label{tbl:bdbr_with_mssim_comparison}
\smallskip\noindent
\resizebox{\linewidth}{!}{%
\setlength\tabcolsep{1.2pt}
\begin{tabular}{@{}ccccccc@{}}
\toprule
\multirow{2}{*}{Class} & \multicolumn{2}{c}{LP}                 & \multicolumn{2}{c}{LD}                 & \multicolumn{2}{c}{RA} \\ \cmidrule(l){2-7} 
                       & Sep.(\%) & Comb.(\%)                 & Sep.(\%) & Comb.(\%)                 & Sep.(\%) & Comb.(\%) \\ \midrule
A                      & -3.90     & \multicolumn{1}{c|}{-3.21} & -2.38     & \multicolumn{1}{c|}{-2.72} & -3.85     & -3.88      \\
B                      & -3.32     & \multicolumn{1}{c|}{-3.24} & -1.67     & \multicolumn{1}{c|}{-1.36} & -1.27     & -1.10       \\
C                      & -3.05     & \multicolumn{1}{c|}{-2.56} & -2.08     & \multicolumn{1}{c|}{-1.50} & -2.51     & -2.16      \\
D                      & -2.79     & \multicolumn{1}{c|}{-2.04} & -2.02     & \multicolumn{1}{c|}{-1.46} & -2.82     & -2.38      \\
E                      & -8.17     & \multicolumn{1}{c|}{-7.30} & -5.73     & \multicolumn{1}{c|}{-4.97} & \multicolumn{2}{c}{-}  \\ \midrule
Average                & -4.02     & \multicolumn{1}{c|}{-3.50} & -2.73     & \multicolumn{1}{c|}{-2.17} & -2.36     & -2.09      \\ \bottomrule
\end{tabular}}
\end{table}

In addition to the PSNR-based BD-Bitrate analysis in Table~\ref{tbl:overall_coding_performance}, we also perform BD-Bitrate analysis with MS-SSIM used instead of PSNR. The results are shown in Table~\ref{tbl:bdbr_with_mssim_comparison}. These results indicate the average bitrate saving of the proposed method compared to HEVC at the same perceptual quality level (as measured by MS-SSIM). The results are qualitatively similar to the results in Table~\ref{tbl:overall_coding_performance}, with savings ranging from 2.1\%-4.0\%. The largest savings are again achieved on sequences in Class E, followed by Classes A and B.  

\begin{figure}[t]
    \centering
    \begin{minipage}[t]{1\linewidth}
    \centering
    \includegraphics[width=\textwidth]{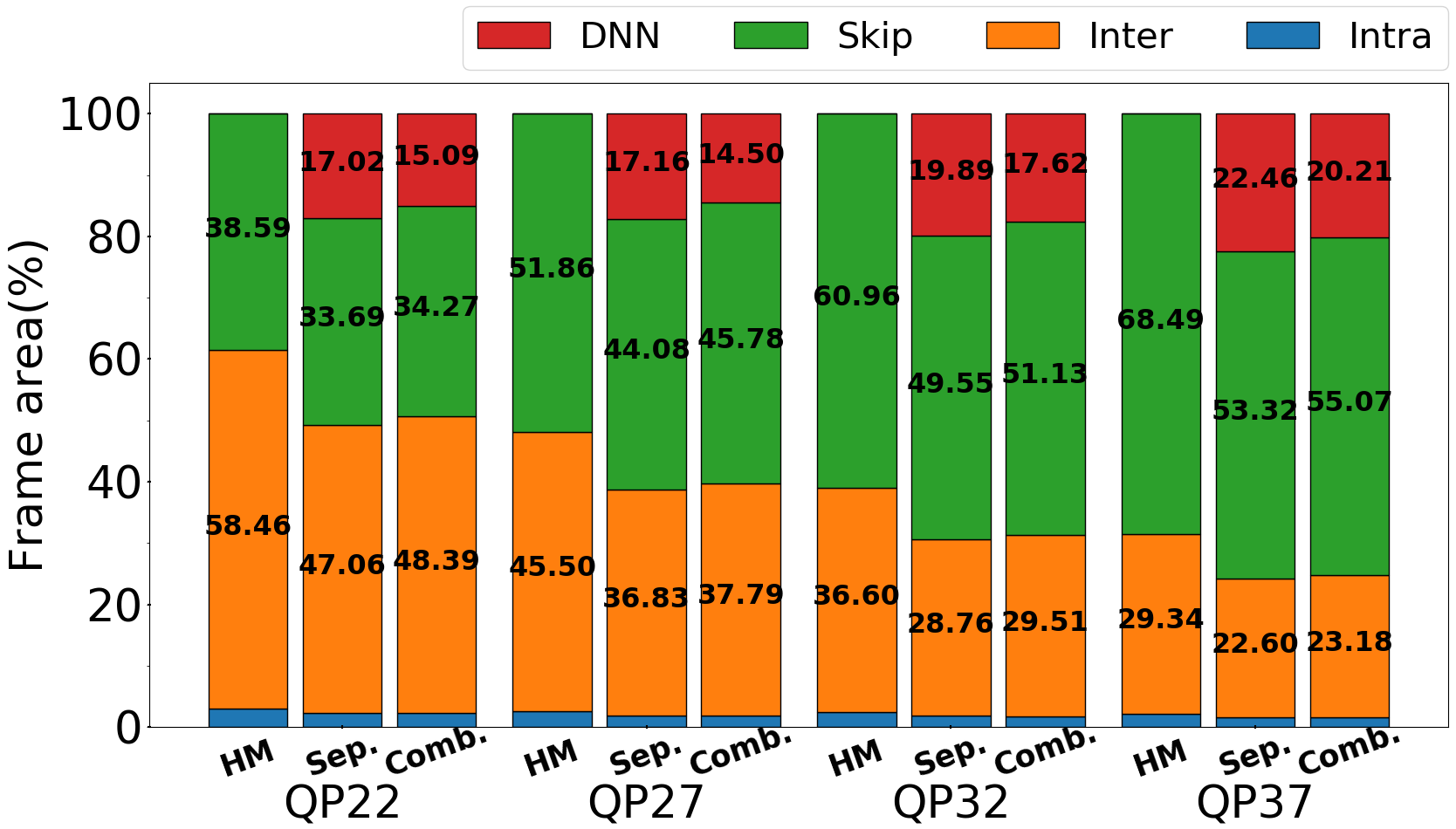}
    \centerline{(a)}
    \end{minipage}
    
    \begin{minipage}[t]{1\linewidth}
    \centering
    \includegraphics[width=\textwidth]{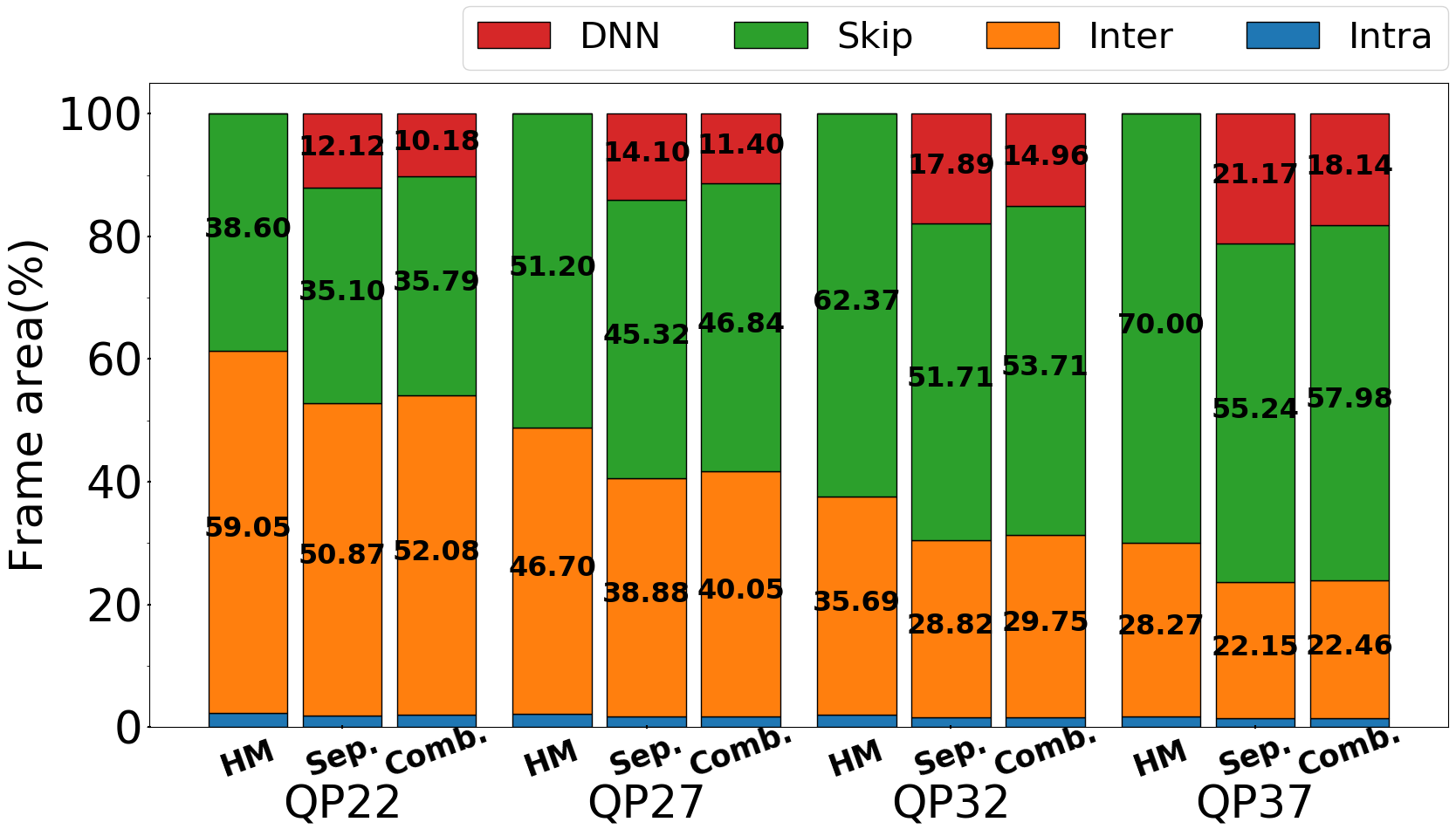}
    \centerline{(b)}
    \end{minipage}
    
    \begin{minipage}[t]{1\linewidth}
    \centering
    \includegraphics[width=\textwidth]{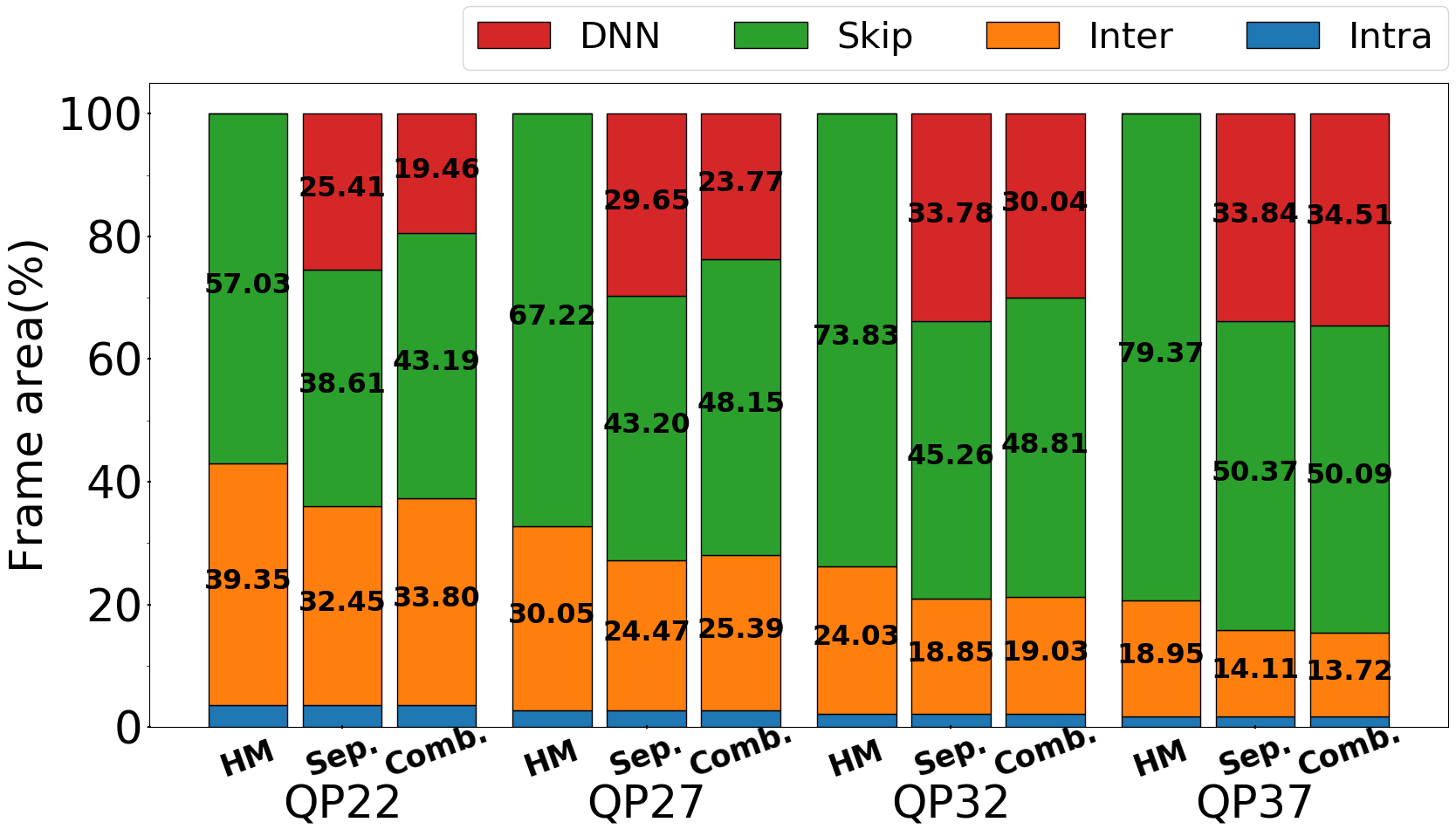}
    \centerline{(c)}
    \end{minipage}
\caption{Selected coding modes for different configurations over various QP values for BQMall in (a) LP, (b) LD, and (c) RA configuration.}
\label{fig:bitstream_analysis}
\end{figure}

\begin{table}[t]
\centering
\caption{Bit reduction percentage in different parts of the bitstream for three configurations with BQMall sequence}
\label{tbl:bitstream_analysis}
\smallskip\noindent
\resizebox{\linewidth}{!}{%
\setlength\tabcolsep{2pt}
\begin{tabular}{@{}ccc|ccccccc@{}}
\toprule
\multirow{2}{*}{Model} & \multirow{2}{*}{Conf.} & \multirow{2}{*}{Sum} & \multicolumn{7}{c}{$\Delta B_{i} (\%)$}                                       \\ \cmidrule(l){4-10} 
                       &                        &                      & $Blk.$    & $DNN$    & $Skip$    & $Inter$   & $Intra$   & $Resi.$   & $SAO$     \\ \midrule
\multirow{3}{*}{Sep.}  & LP                     & -5.00              & -1.56 & 2.63 & -0.44 & -6.23 & -0.83 & 1.69 & -0.26 \\
                       & LD                     & -4.16              & -1.27 & 2.41 & -0.30 & -5.87 & -0.51 & 1.58 & -0.21 \\
                       & RA                     & -3.76              & -0.75 & 0.80 & -0.14 & -3.59 & -0.05 &-0.02 & -0.02 \\ \midrule
\multirow{3}{*}{Comb.} & LP                     & -4.34              & -1.46 & 2.55 & -0.34 & -5.69 & -0.88 & 1.75 & -0.27 \\
                       & LD                     & -3.33              & -1.13 & 2.28 & -0.19 & -5.20 & -0.50 & 1.60 & -0.19 \\
                       & RA                     & -3.44              & -0.73 & 0.76 & -0.11 & -3.33 & -0.04 & 0.03 & -0.02 \\ \bottomrule
\end{tabular}}
\end{table}

\textcolor{red}{}

In Fig.~\ref{fig:bitstream_analysis}, we show the percentage area of the frame where a particular coding mode is selected. In order to compare the coding mode use depending on the coding configuration, Fig.~\ref{fig:bitstream_analysis}(a), (b), and (c) show the case of BQMall sequence in the LP, LD, and RA configuration, respectively. For each QP value, three bars are shown: one for HM-16.20, one for the separate uni- and bi-directional DNNs (``Sep.''), and one for the joint uni-/bi-directional DNN (``Comb.''). From the figures, we see that the proposed frame prediction takes over from inter and skip modes in all cases, but more so in the RA configuration in Fig.~\ref{fig:bitstream_analysis}(c). However, looking at the BD-Bitrate gain for the BQMall sequence in Table~\ref{tbl:overall_coding_performance}, the highest gain is achieved in the LP configuration. This is because, the more inter blocks are taken over by the proposed method, the more gain is achieved, because bits are saved by not transmitting motion vectors and associated information.

Further, we analyze bit savings in different parts of the bitstream. Table~\ref{tbl:bitstream_analysis} shows the average bit reduction percentage in various parts of the bitstream: 
\begin{equation}
\Delta B_{i} = \frac{B_{Proposed, i}-B_{HM, i}}{\sum_{i} B_{HM, i}} \times 100\%
\label{eq:delta_syntax_ratio}
\end{equation}
\noindent where $B_{Proposed, i}$ are the bits used by the proposed method in bitstream part $i$, and $B_{HM, i}$ are the bits used by HM in the same part. The different parts of the bitstream are listed in the second row in the table. $Blk.$ includes split, block size, and prediction mode information. $DNN$ and $Skip$ denote indicator flags for each mode. All motion information and intra prediction directional information are represented by $Inter$ and $Intra$, respectively. $Resi.$ represents bits related to the residual signal and $SAO$ is for sample adaptive offset~\cite{hevc_std}. The third column (``Sum'') in Table~\ref{tbl:bitstream_analysis} shows the sum of bit savings over each row. The highest bit saving is achieved on the $Inter$ part in the LP configuration, specifically 6.23\% and 5.69\% by Sep and  Comb models, respectively. While the proposed methods do use extra bits to signal $DNN$ prediction and sometimes need extra bits for $Resi.$, when aggregated over all parts of the bitstrem, the proposed strategy saves at least 3.3\% of bits compared to HEVC. 
As expected, bit savings by Sep are slightly higher than those achieved by Comb, but Comb model still shows solid savings over all configurations.

\begin{figure}[t]
    \centering
    \begin{minipage}[t]{1\linewidth}
    \centering
    \includegraphics[width=\textwidth]{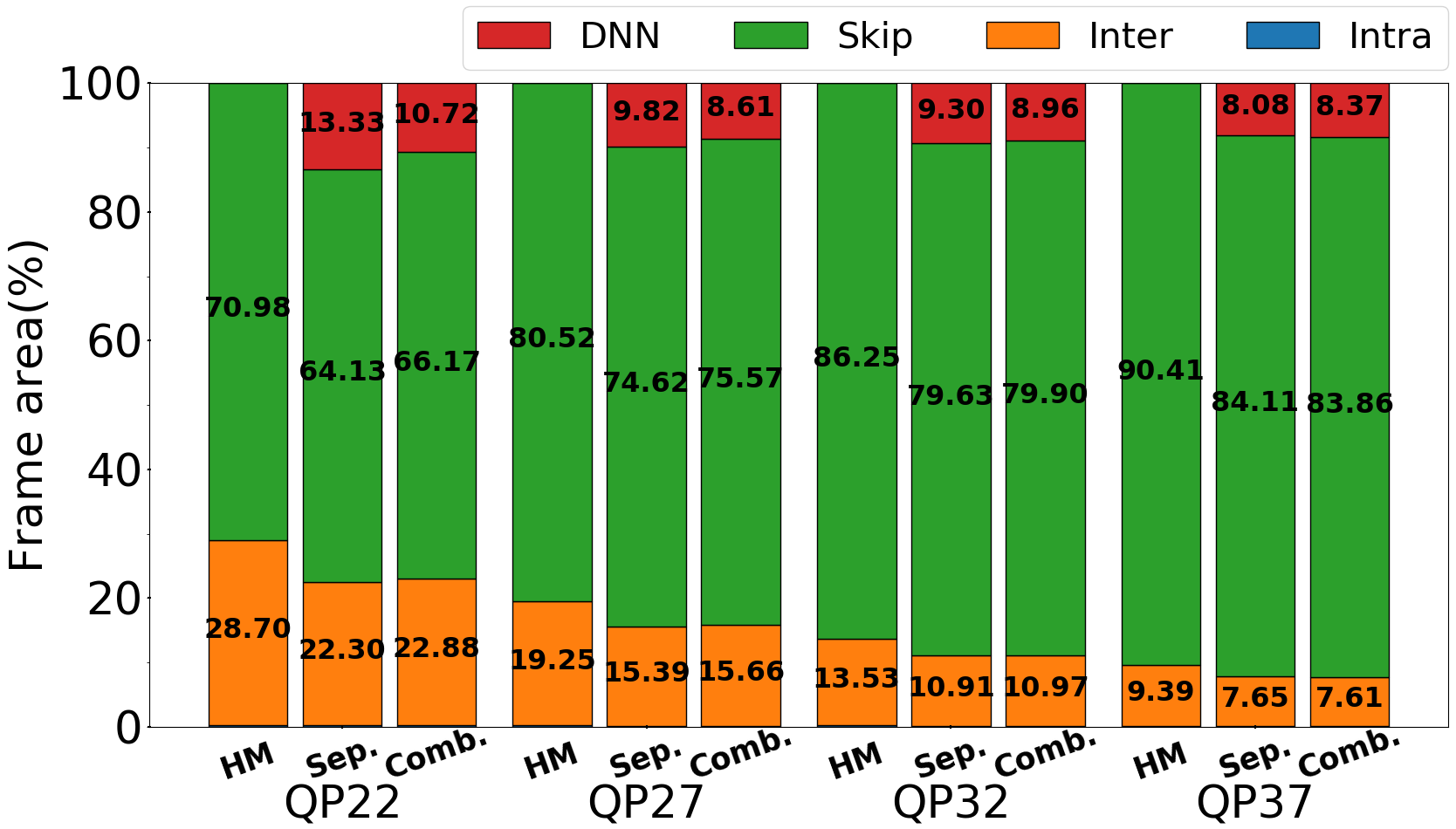}
    \centerline{(a)}
    \end{minipage}
    
    \begin{minipage}[t]{1\linewidth}
    \centering
    \includegraphics[width=\textwidth]{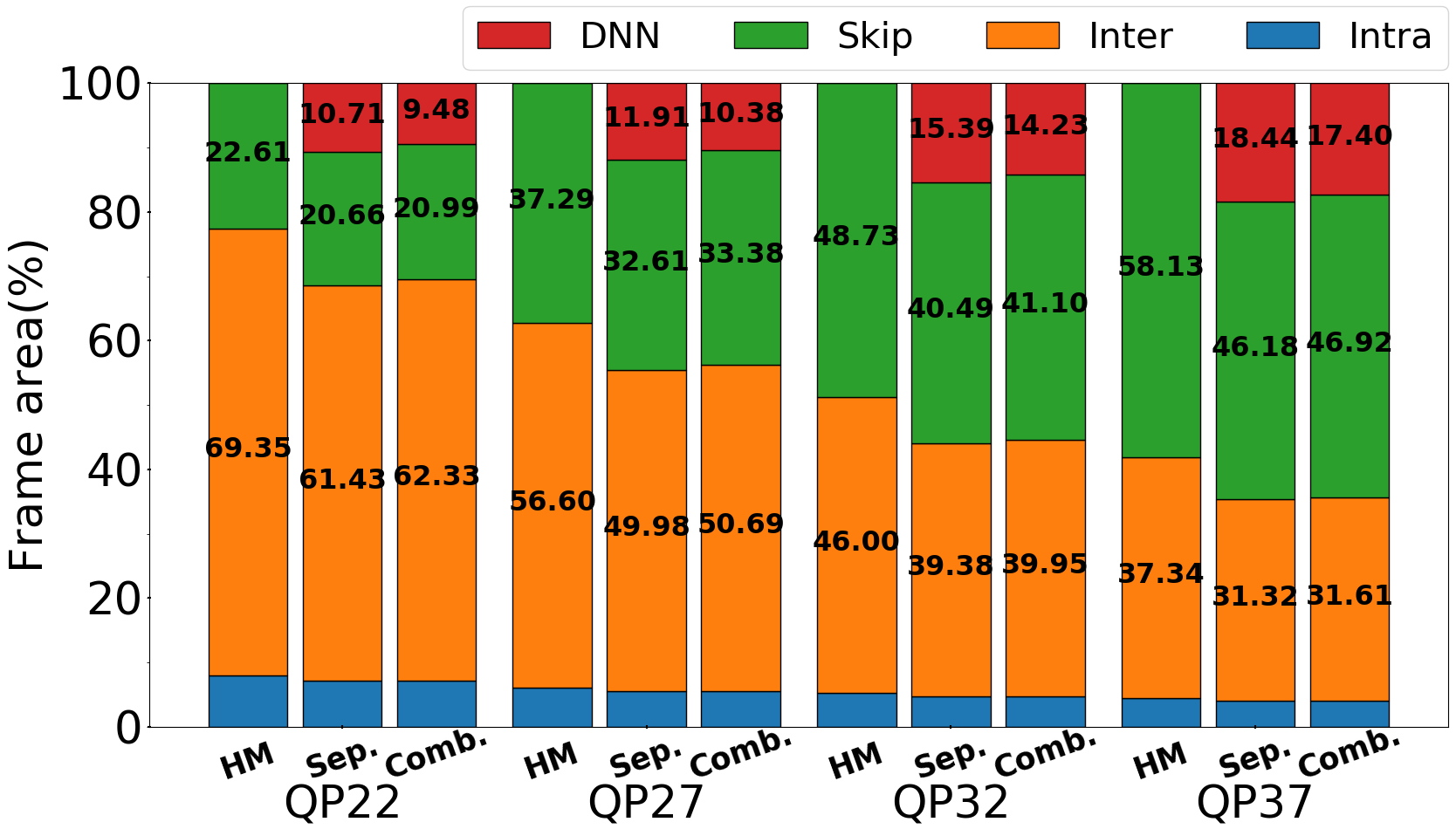}
    \centerline{(b)}
    \end{minipage}
    
\caption{Selected coding modes for (a) Class E and (b) dynamic sequences over various QP values in the LD configuration.}
\label{fig:static_dynamic_bitstream_analysis}
\end{figure}

\begin{table}[t]
\centering
\caption{Bit reduction percentage in different parts of the bitstream for Class E and dynamic sequences}
\label{tbl:static_dynamic_bitstream_analysis}
\smallskip\noindent
\resizebox{\linewidth}{!}{%
\setlength\tabcolsep{2pt}
\begin{tabular}{@{}ccc|ccccccc@{}}
\toprule
\multirow{2}{*}{Group}   & \multirow{2}{*}{Model} & \multirow{2}{*}{Sum} & \multicolumn{7}{c}{$\Delta B_{i} (\%)$}                                       \\ \cmidrule(l){4-10} 
                         &                        &                      & $Blk.$    & $DNN$    & $Skip$    & $Inter$   & $Intra$   & $Resi.$   & $SAO$     \\ \midrule
\multirow{2}{*}{Class E}  & Sep.                    & -4.42              & -1.67 & 2.62 & 0.24  & -6.28 & -0.20 & 1.25 & -0.38 \\
                         & Comb.                   & -3.74              & -1.62 & 2.59 & 0.27  & -6.11 & -0.20 & 1.72 & -0.38 \\ \midrule
\multirow{2}{*}{Dynamic} & Sep.                    & -1.42              & -0.50 & 1.66 & -0.12 & -2.91 & -0.23 & 0.78 & -0.09 \\
                         & Comb.                   & -1.20              & -0.48 & 1.59 & -0.12 & -2.75 & -0.22 & 0.84 & -0.07 \\ \bottomrule
\end{tabular}}
\end{table}

To further investigate the reasons behind larger coding gains in Class E in Table~\ref{tbl:overall_coding_performance}, we first note that Class E sequences are fairly static, consisting of static background and slowly-moving objects. We therefore select a group of more dynamic sequences for comparison: BasketballDrive, RaceHorsesC, BlowingBubbles. These dynamic sequences contain camera motion/moving background, as well as faster-moving objects.  
The average BD-Bitrate from  Table~\ref{tbl:overall_coding_performance} in the LD configuration is $-$5.43\% for Class E and $-$1.38\% for the three dynamic sequences.
Fig.~\ref{fig:static_dynamic_bitstream_analysis} shows selected coding modes in each group. In Class E sequences, (Fig.~\ref{fig:static_dynamic_bitstream_analysis}(a)), large static background causes skip mode to become more efficient as QP increases, while the proposed method and inter prediction are getting less selected. Meanwhile, in dynamic sequences  (Fig.~\ref{fig:static_dynamic_bitstream_analysis}(b)), both our predictor and skip mode tend to win over a larger area as QP increases, at the expense of inter prediction. The area over which our method is used at QP=37 in dynamic sequences is twice as large as the corresponding area in Class E sequences (17.4-18.4\% vs. 8.1-8.4\%). Still, larger area does not translate into larger coding gain, because our gain is actually larger in Class E sequences. This is due to the fact that the bitrate for Class E sequences is relatively small to start with, due to the very frequent use of the skip mode, so even a slight reduction in absolute bits could represent a larger relative saving in such sequences. 
Further, Table~\ref{tbl:static_dynamic_bitstream_analysis} shows bit reduction percentage in different parts of the bitstream in the two groups of sequences. As seen in the table, the larger bit reduction in Class E sequences is mostly due to the much reduced bitrate for $Inter$ prediction.

Finally, we compare the coding efficiency of the proposed method to that of Lei \textit{et al.}~\cite{hevc_with_sep_conv_for_ra}, which directly uses the network from~\cite{Niklaus_ICCV_2017}. Since the borrowed network was designed for interpolation mid-way between two frames,~\cite{hevc_with_sep_conv_for_ra} could only test and provide results for the RA configuration. Hence, we carry out this comparison on the RA configuration only. Following the experimental setup in~\cite{hevc_with_sep_conv_for_ra}, we re-implemented our proposed method in HM-16.6, and evaluated the coding performance on sequences in Class B, C and D for 2 seconds with QP $\in \{27, 32, 37, 42\}$. Table~\ref{tbl:performance_comparison} shows BD-Bitrate of each method against HM-16.6. Overall, our separately trained DNN model for bi-directional prediction (``Sep.'') is the best overall, with the highest average reduction in bitrate (3.3\%) and providing the best performance in 9 out of 13 sequences in this test. Lei \textit{et al.} results are the second best overall, with the average bit rate reduction of 3.2\% and top performance in 4 out of 13 sequences. They achieve especially good performance on BQSquare, which significantly boosts their average bit saving. Our combined uni-/bi-directional DNN (``Comb.'') comes in third with a slightly lower overall bit rate reduction of 3.1\%. However, even our combined DNN provides better coding gain than Lei \textit{et al.} in 8 out of 13 sequences. Furthermore, despite the fact that DNN-based uni-directional prediction was not used in this test, it is worth remembering that our combined DNN is the only one of the three models that supports both uni- and bi-directional prediction, and still is competitive with the other two models.

\begin{table}[t]
\centering
\caption{BD-Bitrate relative to HM-16.6 with the RA configuration}
\label{tbl:performance_comparison}
\smallskip\noindent
\resizebox{\linewidth}{!}{%
\setlength\tabcolsep{2pt}
\begin{tabular}{@{}ccccc|cc|cc@{}}
\toprule
\multirow{3}{*}{Class} & \multirow{3}{*}{Sequence} & \multirow{3}{*}{fps} & \multicolumn{2}{c}{\multirow{2}{*}{Lei \textit{et al.}\cite{hevc_with_sep_conv_for_ra}}} & \multicolumn{4}{c}{Proposed}                                    \\ \cmidrule(l){6-9} 
                       &                            &                      & \multicolumn{2}{c}{}                            & \multicolumn{2}{c}{Sep.}       & \multicolumn{2}{c}{Comb.}      \\ \cmidrule(l){4-9} 
                       &                            &                      & Y (\%)           & Avg. (\%)                    & Y (\%)                & Avg. (\%)              & Y (\%)             & Avg. (\%)             \\ \midrule
\multirow{5}{*}{B}     & BQTerrace                  & 60                   & --0.2            & \multirow{5}{*}{--2.0}       & \textbf{--1.0}        & \multirow{5}{*}{--1.9} &         --0.7      & \multirow{5}{*}{--1.9}     \\
                       & BasketballDrive            & \multirow{2}{*}{50}  & --1.1            &                              & \textbf{--1.5}        &                        &         --1.4      &                       \\
                       & Cactus                     &                      & \textbf{--4.6}   &                              &         --3.1         &                        &         --3.6      &                       \\
                       & Kimono                     & \multirow{2}{*}{24}  & \textbf{--1.7}   &                              &         --0.9         &                        &         --0.9      &                       \\
                       & ParkScene                  &                      & --2.6            &                              & \textbf{--2.9}        &                        & \textbf{--2.9}     &                       \\ \midrule
\multirow{4}{*}{C}     & BQMall                     & 60                   & --6.0            & \multirow{4}{*}{--3.2}       & \textbf{--7.1}        & \multirow{4}{*}{--3.8} &         --6.2      & \multirow{4}{*}{--3.5} \\
                       & BasketballDrill            & \multirow{2}{*}{50}  & \textbf{--3.2}   &                              &         --2.4         &                        &         --2.6      &                       \\
                       & PartyScene                 &                      & --3.0            &                              & \textbf{--4.3}        &                        &         --3.9      &                       \\
                       & RaceHorsesC                & 30                   & --0.8            &                              & \textbf{--1.4}        &                        &         --1.3      &                       \\ \midrule
\multirow{4}{*}{D}     & BQSquare                   & 60                   & \textbf{--7.1}   & \multirow{4}{*}{--4.7}       &         --4.7         & \multirow{4}{*}{--4.6} &         --4.2      & \multirow{4}{*}{--4.3} \\
                       & BasketballPass             & \multirow{2}{*}{50}  & --5.4            &                              & \textbf{--6.7}        &                        &         --6.2      &                       \\
                       & BlowingBubbles             &                      & --4.1            &                              & \textbf{--4.3}        &                        &         --3.9      &                       \\
                       & RaceHorses                 & 30                   & --2.2            &                              & \textbf{--2.9}        &                        & \textbf{--2.9}     &                       \\ \midrule
\multicolumn{3}{c}{Average}                                                & \multicolumn{2}{c}{--3.2}                       &  \multicolumn{2}{c}{\textbf{--3.3}}      &  \multicolumn{2}{c}{--3.1}                       \\ \bottomrule
\end{tabular}}
\end{table}

\section{Conclusion}
\label{sec:conclusion}
We presented a deep neural network (DNN) for frame prediction that can be used to improve video coding efficiency. The DNN operates at both encoder and decoder, and uses previously decoded frames to predict the current frame. This form of prediction is signalled as a separate prediction mode, and can be used within RD optimization to compete with other prediction modes. Although it is a form of inter prediction, it does not require any motion vectors to be transmitted. Three DNNs were trained for this purpose: two for separate uni- and bi-directional prediction, and one that supports both uni- and bi-directional prediction, the first of its kind, to our knowledge. The DNNs were evaluated on common test sequences and various coding configurations, and demonstrated to bring significant coding gains relative to HEVC. Furthermore, advantages over the recent work on DNN-based frame prediction for video coding was also demonstrated.

\ifCLASSOPTIONcaptionsoff
  \newpage
\fi

\bibliographystyle{IEEEbib}
\bibliography{ref}

\begin{IEEEbiography}
    [{\includegraphics[width=1in,height=1.25in,clip,keepaspectratio]{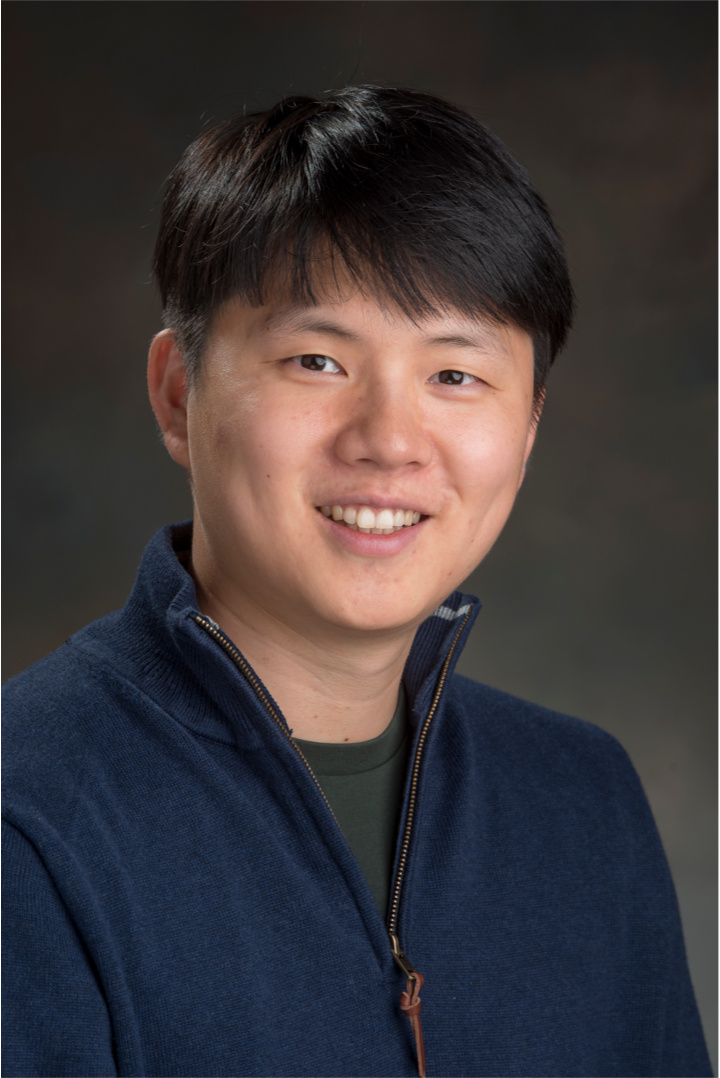}}]{Hyomin Choi} (S'17) is a Ph.D student under the supervision of Prof. Ivan V. Baji\'{c} at Simon Fraser University, Burnaby, BC, Canada. He received B.S. and M.S degrees in computer engineering from Kwangwoon University, Seoul, Korea, in 2010 and 2012, respectively. He was a research engineer at System IC Research Center, LG Electronics from 2012 to 2016.  
    His research interests include image/video coding and deep learning. He is the winner of the Vanier Canada Graduate Scholarships (Vanier CGS) in 2017. He was the recipient of the IEEE SPS Student Travel Grant for ICIP 2018.
\end{IEEEbiography}

\begin{IEEEbiography}
    [{\includegraphics[width=1in,height=1.25in,clip,keepaspectratio]{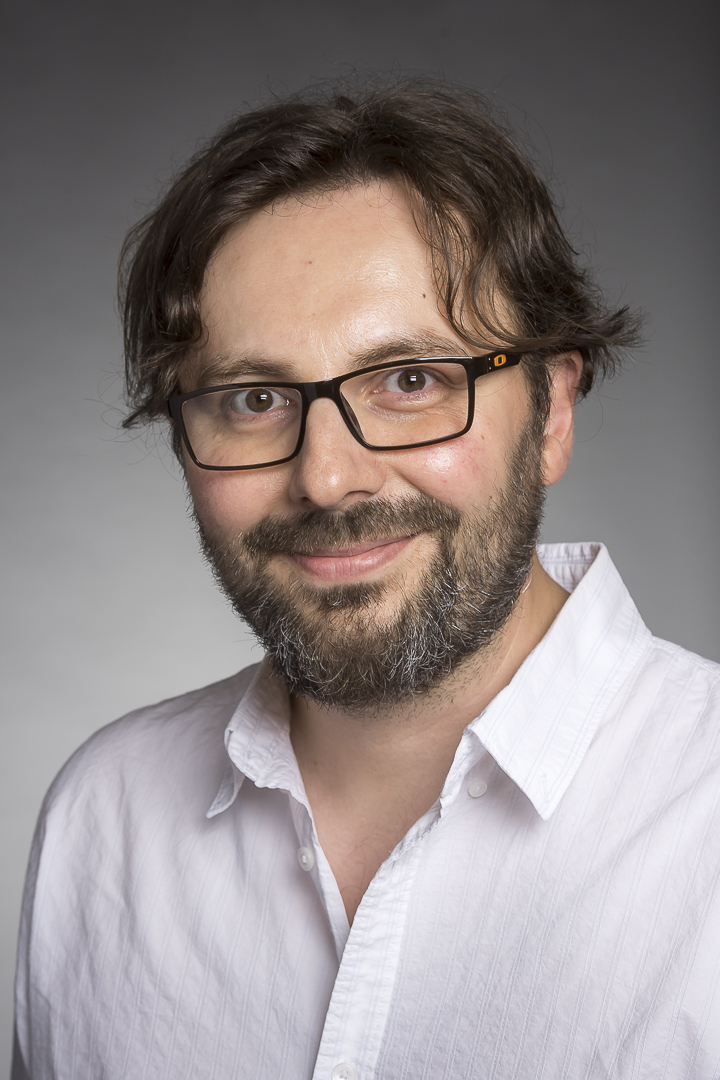}}]{Ivan V. Baji\'{c}} (S'99-M'04-SM'11) is Professor of Engineering Science and co-director of the Multimedia Lab at Simon Fraser University, Burnaby, BC, Canada. His research interests include signal processing and machine learning with applications to multimedia processing, ergonomics, compression, and communications. He has authored about a dozen and co-authored another ten dozen publications in these fields. He was the Chair of the Media Streaming Interest Group of the IEEE Multimedia Communications Technical Committee from 2010 to 2012, and is currently an elected member of the IEEE Multimedia Signal Processing Technical Committee and the IEEE Multimedia Systems and Applications Technical Committee. He has served on the organizing and/or program committees of the main conferences in the field. He was an Associate Editor of \textsc{IEEE Transactions on Multimedia} and \textsc{IEEE Signal Processing Magazine}, and is currently an Area Editor of \textsc{Signal Processing: Image Communication}. He was the Chair of the IEEE Signal Processing Society Vancouver Chapter for several years, during which the Chapter received the Chapter of the Year Award from IEEE SPS. 
\end{IEEEbiography}

\end{document}